\documentclass[twocolumn,trackchanges]{aastex62}
\usepackage{bm}
\usepackage{ulem}
\usepackage{comment}
\usepackage{longtable}

\received{MMM DD, YYYY}
\revised{MMM DD, YYYY}
\accepted{MMM DD, YYYY}
\submitjournal{AJ}

\shorttitle{SSA22-HIT-catalog}
\shortauthors{Mawatari et al.}

\begin{document}

\title{The SSA22 H\,{\sc i} Tomography Survey (SSA22-HIT). I. Data Set and Compiled Redshift Catalog}

\correspondingauthor{Ken Mawatari}
\email{ken.mawatari@nao.ac.jp}

\author[0000-0003-4985-0201]{Ken Mawatari}
\affiliation{National Astronomical Observatory of Japan, Osawa 2-21-1, Mitaka, Tokyo 181-8588, Japan}
\affiliation{Institute for Cosmic Ray Research, The University of Tokyo, 5-1-5 Kashiwanoha, Kashiwa, Chiba 277-8582, Japan}
\affiliation{Department of Environmental Science and Technology, Fuculty of Design Technology, Osaka Sangyo University, 3-1-1, Nakagaito, Daito, Osaka, 574-8530, Japan}

\author[0000-0002-7779-8677]{Akio K. Inoue}
\affiliation{Department of Physics, School of Advanced Science and Engineering, Faculty of Science and Engineering, Waseda University, 3-4-1, Okubo, Shinjuku, Tokyo 169-8555, Japan}
\affiliation{Waseda Research Institute for Science and Engineering, Faculty of Science and Engineering, Waseda University, 3-4-1, Okubo, Shinjuku, Tokyo 169-8555, Japan}
\affiliation{Department of Environmental Science and Technology, Faculty of Design Technology, Osaka Sangyo University, 3-1-1, Nakagaito, Daito, Osaka, 574-8530, Japan}

\author{Toru Yamada}
\affiliation{Institute of Space Astronautical Science, Japan Aerospace Exploration Agency, Sagamihara, Kanagawa 252-5210, Japan}
\affiliation{Astronomical Institute, Tohoku University, Aoba, Aramaki, Aoba-ku, Sendai, Miyagi, 980-8578, Japan}

\author{Tomoki Hayashino}
\affiliation{Research Center for Neutrino Science, Graduate School of Science, Tohoku University, Aoba, Aramaki, Aoba-ku, Sendai, Miyagi, 980-8578, Japan}

\author[0000-0002-7738-6875]{J. Xavier Prochaska}
\affiliation{University of California Observatories, Lick Observatory, 1156 High Street, Santa Cruz, CA 95064, USA}
\affiliation{Department of Astronomy and Astrophysics, University of California at Santa Cruz, 1156 High Street, Santa Cruz, CA 95064, USA}
\affiliation{Kavli Institute for the Physics and Mathematics of the Universe, The University of Tokyo, 5-1-5 Kashiwanoha, Kashiwa, Chiba 277-8583, Japan}

\author[0000-0001-9299-5719]{Khee-Gan Lee}
\affiliation{Kavli Institute for the Physics and Mathematics of the Universe, The University of Tokyo, 5-1-5 Kashiwanoha, Kashiwa, Chiba 277-8583, Japan}
\affiliation{Lawrence Berkeley National Laboratory, 1 Cyclotron Road, Berkeley, CA 94720, USA}

\author[0000-0002-1883-4252]{Nicolas Tejos}
\affiliation{Instituto de F{\'i}sica, Pontificia Universidad Cat{\'o}lica de Valpara{\'i}so, Casilla 4059, Valpara{\'i}so, Chile}
\affiliation{University of California Observatories, Lick Observatory, 1156 High Street, Santa Cruz, CA 95064, USA}

\author[0000-0001-5493-6259]{Nobunari Kashikawa}
\affiliation{Department of Astronomy, Graduate School of Science, The University of Tokyo, 7-3-1 Hongo, Bunkyo, Tokyo 113-0033, Japan}
\affiliation{National Astronomical Observatory of Japan, Osawa 2-21-1, Mitaka, Tokyo 181-8588, Japan}
\affiliation{The Graduate University for Advanced Studies (SOKENDAI), Tokyo 181-8588, Japan}

\author{Takuya Otsuka}
\affiliation{Astronomical Institute, Tohoku University, Aoba, Aramaki, Aoba-ku, Sendai, Miyagi, 980-8578, Japan}

\author[0000-0002-7738-5290]{Satoshi Yamanaka}
\affiliation{General Education Department, National Institute of Technology, Toba College, 1-1, Ikegami-cho, Toba, Mie 517-8501, Japan}
\affiliation{Research Center for Space and Cosmic Evolution, Ehime University, 2-5, Bunkyo-cho, Matsuyama, Ehime 790-8577, Japan}
\affiliation{Waseda Research Institute for Science and Engineering, Faculty of Science and Engineering, Waseda University, 3-4-1, Okubo, Shinjuku, Tokyo 169-8555, Japan}
\affiliation{Department of Environmental Science and Technology, Fuculty of Design Technology, Osaka Sangyo University, 3-1-1, Nakagaito, Daito, Osaka, 574-8530, Japan}

\author[0000-0002-5042-5088]{David J. Schlegel}
\affiliation{Lawrence Berkeley National Laboratory, 1 Cyclotron Road, Berkeley, CA 94720, USA}

\author{Yuichi Matsuda}
\affiliation{National Astronomical Observatory of Japan, Osawa 2-21-1, Mitaka, Tokyo 181-8588, Japan}

\author[0000-0002-7054-4332]{Joseph F. Hennawi}
\affiliation{Department of Physics, Broida Hall, University of California at Santa Barbara, Santa Barbara, CA 93106, USA}
\affiliation{Max Planck Institute for Astronomy, Ko{\''n}igstuhl 17, D-69117 Heidelberg, Germany}

\author{Ikuru Iwata}
\affiliation{National Astronomical Observatory of Japan, Osawa 2-21-1, Mitaka, Tokyo 181-8588, Japan}
\affiliation{The Graduate University for Advanced Studies (SOKENDAI), Tokyo 181-8588, Japan}

\author[0000-0003-1937-0573]{Hideki Umehata}
\affiliation{Institute of Advanced Research, Graduate School of Science, Nagoya University, Furo-cho, Chikusa-ku, Nagoya, Aichi 464-8601, Japan}
\affiliation{Department of Physics, Graduate School of Science, Nagoya University, Furocho, Chikusa, Nagoya 464-8602, Japan}
\affiliation{Institute for Cosmic Ray Research, The University of Tokyo, 5-1-5 Kashiwanoha, Kashiwa, Chiba 277-8582, Japan}

\author{Shiro Mukae}
\affiliation{Institute for Cosmic Ray Research, The University of Tokyo, 5-1-5 Kashiwanoha, Kashiwa, Chiba 277-8582, Japan}
\affiliation{Department of Astronomy, Graduate School of Science, The University of Tokyo, 7-3-1 Hongo, Bunkyo, Tokyo, 113-0033, Japan}

\author[0000-0002-1049-6658]{Masami Ouchi}
\affiliation{Institute for Cosmic Ray Research, The University of Tokyo, 5-1-5 Kashiwanoha, Kashiwa, Chiba 277-8582, Japan}
\affiliation{Division of Science, National Astronomical Observatory of Japan, 2-21-1 Osawa, Mitaka, Tokyo 181-8588, Japan}

\author[0000-0001-6958-7856]{Yuma Sugahara}
\affiliation{Waseda Research Institute for Science and Engineering, Faculty of Science and Engineering, Waseda University, 3-4-1, Okubo, Shinjuku, Tokyo 169-8555, Japan}
\affiliation{National Astronomical Observatory of Japan, Osawa 2-21-1, Mitaka, Tokyo 181-8588, Japan}
\affiliation{Institute for Cosmic Ray Research, The University of Tokyo, 5-1-5 Kashiwanoha, Kashiwa, Chiba 277-8582, Japan}

\author[0000-0003-4807-8117]{Yoichi Tamura}
\affiliation{Division of Particle and Astrophysical Science, Graduate School of Science, Nagoya University, Nagoya 464-8602, Japan}
\affiliation{Institute of Astronomy, The University of Tokyo, 2-21-1 Osawa, Mitaka, Tokyo 181-0015, Japan}

\begin{abstract}
We conducted a deep spectroscopic survey, named SSA22-HIT, in the SSA22 field with the DEep Imaging MultiObject Spectrograph (DEIMOS) on the Keck telescope, designed to tomographically map high-$z$ H\,{\sc i} gas through analysis of Ly$\alpha$ absorption in background galaxies' spectra. In total, 198 galaxies were spectroscopically confirmed at $2.5 < z < 6$ with a few low-$z$ exceptions in the $26 \times 15$\,arcmin$^2$ area, of which 148 were newly determined in this study. 
Our redshift measurements were merged with previously confirmed redshifts available in the $34 \times 27$\,arcmin$^2$ area of the SSA22 field. This compiled catalog containing $730$ galaxies of various types at $z > 2$ is useful for various applications, and it is made publicly available. 
Our SSA22-HIT survey has increased by approximately twice the number of spectroscopic redshifts of sources at $z > 3.2$ in the observed field. 
From a comparison with publicly available redshift catalogs, we show that our compiled redshift catalog in the SSA22 field is comparable to those among major extragalactic survey fields in terms of a combination of wide area and high surface number density of objects at $z > 2$. 
About 40\,\% of the spectroscopically confirmed objects in SSA22-HIT show reasonable quality of spectra in the wavelengths shorter than Ly$\alpha$ when a sufficient amount of smoothing is adopted. Our data set enables us to make the H\,{\sc i} tomographic map at $z \gtrsim 3$, which we present in a parallel study.

\end{abstract}



\section{Introduction} \label{sec:intro}

It is important to investigate how galaxies are formed through the assembly of gaseous matter.
Inter-/circumgalactic medium (IGM/CGM) in the high-redshift universe was probed using the H\,{\sc i} Ly$\alpha$ and metal absorption lines imprinted in the spectra of background QSOs (\citealt{Rauch+98,Adelberger+03,Peroux+03,Wolfe+05,Hennawi+06,Rakic+12,Rudie+12,Prochaska+14,Turner+14,Boksenberg+15,Crighton+15,Cai+16,Mukae+17}). In the past two decades, star-forming galaxies have also been used as background light sources \citep[e.g.][]{Adelberger+05,Rubin+10,Steidel+10,CookeOMeara15,Mawatari+16a,Lopez+18,Peroux+18}. 
While the absorption can be measured only along sight-lines of background light sources, the higher surface number density of star-forming galaxies than those of QSOs makes it possible to interpolate between the discrete sight-lines and to reconstruct the three-dimensional IGM large-scale structures. In the past several years, a novel approach to resolve H\,{\sc i} gas structures on megaparsec scales using a Wiener-filtered tomographic method was introduced by the COSMOS Ly$\alpha$ Mapping and Tomography Observations project (CLAMATO; \citealt{KGLee+14b,KGLee+16,KGLee+18,Krolewski+18,Horowitz+22}). This ``IGM H\,{\sc i} tomography'' technique has been successfully applied in subsequent works \citep{Mukae+20a,Mukae+20b,Newman+20,Ravoux+20}.  

The spatial resolutions of the H\,{\sc i} tomography in the celestial plane and along the line of sight are determined by the surface number density of the background light sources and the spectral resolution of the observations, respectively. These are limited to a few comoving megaparsecs (cMpc), even with the deepest observations of $z = 2$--$3$ galaxies \citep{KGLee+18,Horowitz+22}. The current spatial resolution is suitable to probe IGM large-scale structures, often called the ``Cosmic Web'' (e.g., \citealt{Mo+10} and references therein). Future high-sensitivity instruments such as the Wide Field Optical Spectrograph (WFOS) on the Thirty Meter Telescope \citep{Pazder+06} and MOSAIC on the European Extremely Large Telescope \citep{Hammer+14,Puech+18} will enable finer-resolution H\,{\sc i} tomography \citep{Pazder+06,Japelj+19} and mapping of metal absorption lines \citep{Thronson+09,Skidmore+15}. 

For H\,{\sc i} tomography, we must detect the spectral continua of individual background galaxies significantly at wavelengths shorter than the galaxies' Ly$\alpha$ line ($\lambda < 1215.67$\,\AA\ in the rest frame). This requirement is more observationally demanding than the detection of emission lines or spectral breaks imposed in many spectroscopic surveys for redshift confirmation. 
At the same time, wide-area observations are also needed to cover large-scale structures over tens of cMpc scales. Many of the previous tomography studies were based on their own fine-tuned observations \citep{KGLee+14b,KGLee+18,Mukae+20a,Newman+20}. 

In this article, we focus particularly on the SSA22 field, which contains one of the largest spatially observed protoclusters \citep{Yamada+12a,Overzier+16,Cai+17a}. The protocluster at $z = 3.1$ in the SSA22 field was initially discovered by \citet{Steidel+98} through a large spectroscopic survey of Lyman break galaxies (LBGs). Several subsequent observations \citep{Hayashino+04,Matsuda+04,Lehmer+09b,Tamura+09,Matsuda+11,Yamada+12a,Uchimoto+12,Kubo+13,Umehata+14,Kubo+15,Umehata+15,Kato+16,Topping+16,Umehata+17b,Umehata+18} have identified this protocluster as a prominent overdensity of a wide variety of objects such as Ly$\alpha$ emitters (LAEs), Ly$\alpha$ blobs (LABs), distant red galaxies (DRGs), submillimeter galaxies (SMGs), and X-ray detected active galactic nuclei (AGNs). High-sensitivity integral field unit (IFU) spectroscopy with the MultiUnit Spectroscopic Explorer \citep{Bacon+10} on the Very Large Telescope (VLT) revealed that the filamentary gas structure traced by the Ly$\alpha$ emission extends over $\sim 4$\,cMpc with a spatial resolution as high as 0.12\,cMpc in the $z = 3.1$ protocluster \citep{Umehata+19}. In addition, a large H\,{\sc i} gas reservoir in/around the protocluster was suggested by \citet{Hayashino+19} and \citet{Mawatari+17}, who detected strong Ly$\alpha$ absorption at $z = 3.1$ in the stacked background and narrowband photometry, respectively. It is desirable to resolve the H\,{\sc i} large-scale structure three-dimensionally and compare it with the spatial distributions of the galaxies and the Ly$\alpha$ emitting gas. 

While a large number of spectroscopic surveys have been conducted in the SSA22 field, many of them aim to confirm the redshift of $z = 3.1$ protocluster member galaxies \citep{Matsuda+05,Matsuda+06,Yamada+12b,Nestor+13,Erb+14,Kubo+15,Kubo+16,Umehata+19}. For H\,{\sc i} tomography in the $z = 3.1$ protocluster, we need to increase the number of background galaxies at $3.2 \lesssim z \lesssim 3.7$, for which deep spectra at wavelengths shorter than Ly$\alpha$ are available. For this reason, we carried out new spectroscopic observations in the SSA22 field using the DEep Imaging MultiObject Spectrograph (DEIMOS; \citealt{Faber+03}) equipped on the Keck telescope. We call this new spectroscopic survey the SSA22 H\,{\sc i} Tomography (SSA22-HIT) survey. 

This is the first study to investigate the H\,{\sc i} distribution in the SSA22 field. In this paper (Paper~I), we focus on the observational aspects of our project, while discussion based on the H\,{\sc i} tomography analysis is described in a companion paper (K. Mawatari et al. 2023, in preparation, hereafter Paper~II). We describe the SSA22-HIT survey design and data reduction in Sections ~\ref{sec:survey_design} and \ref{sec:datareduc}, respectively. The redshift determination in the SSA22-HIT survey and compilation of the archival redshift catalogs are described in Section~\ref{sec:redshift}. In Section~\ref{sec:comp_survey}, we briefly compare our compiled redshift catalog in the SSA22 field with public catalogs available in other extragalactic survey fields. Quality assessment of the individual spectra and discussion for H\,{\sc i} tomography are described in Section~\ref{sec:HITquality}. In this study, we use the AB magnitude system \citep{OkeGunn83} and adopt a cosmology with $H_{0}=70$ km s$^{-1}$ Mpc$^{-1}$, $\Omega_{M}=0.3$, and $\Omega_{\Lambda}=0.7$.

\section{Survey Design}\label{sec:survey_design}

\subsection{Photometric Data Set}\label{sec:imaging_data}
In this study, we focus on the $\sim 34 \times 27$\,arcmin$^2$ region centered at the LAE density peak, referred to as the SSA22-Sb1 field \citep{Hayashino+04,Yamada+12a,Mawatari+17}. We use the following multiband imaging data: Canada France Hawaii Telescope/Megacam \citep{Boulade+03} $u^{*}$-band image \citep{Kousai11}; Subaru/Suprime-Cam \citep{Miyazaki+02} $B$, $V$, $R_{c}$, $i'$, $z'$, NB359 {(central wavelength $\lambda_{\rm cen} = 3596$\,\AA\ and FWHM $= 152$\,\AA)}, NB497 {($\lambda_{\rm cen} = 4977$\,\AA\ and FWHM $= 78$\,\AA)}, and NB816 {($\lambda_{\rm cen} = 8150$\,\AA\ and FWHM $= 118$\,\AA)} band images \citep{Hayashino+04,Iwata+09,Nakamura+11,Yamada+12a}; Subaru/Hyper Suprime-Cam (HSC; \citealt{Miyazaki+12,Miyazaki+18}) $Y$-band image from the Subaru strategic program with HSC (HSC-SSP; \citealt{Aihara+18a, Furusawa+18, Kawanomoto+18, Komiyama+18, Miyazaki+18}) in its internal data products (S14A0b) generated from raw data taken in 2014; Subaru/Multi-Objects Infra-Red Camera and Spectrograph (MOIRCS; \citealt{Ichikawa+06}) $J$-, $H$-, and $K_{s}$-band images \citep{Uchimoto+12}; and Spitzer/IRAC \citep{Fazio+04} $3.6 \mu$m and $4.5 \mu$m band images. The entire SSA22-Sb1 field is covered by the above imaging data, except for the MOIRCS and IRAC data. 

We also created a composite image of the $B$ and $V$ bands (hereafter, the $BV$ image) as $BV = (2 B + V)/3$ following \citet{Yamada+12a} to estimate the continuum flux at the NB497 wavelength. The Suprime-Cam NB816 raw images captured  by \citet{Hu+04} were reduced by us using the pipeline ``SDFRED'' \citep{Yagi+02,Ouchi+04}. For the IRAC images, we collected raw images via the NASA/IPAC Infrared Science Archive and reduced them using a standard pipeline MOPEX\footnote{\url{https://irsa.ipac.caltech.edu/data/SPITZER/docs/dataanalysistools/tools/mopex/}}. Since astrometric correction in the used images was applied independently by different authors, we again calibrated astrometry to that of the stars in the USNO-B1.0 catalog \citep{Monet+03} using NOAO/IRAF\footnote{\url{https://iraf-community.github.io/}}. The resultant astrometric uncertainties are typically less than 1\arcsec, whereas the relative offsets among the different band images are much smaller. 

Object extraction was performed on the $i'$- and $z'$-band images using SExtractor \citep{BertinArnouts96} version 2.5.0. We constructed multiband photometry catalogs for objects detected in the $i'$ and $z'$ bands, adopting 2.2\arcsec diameter apertures in all band photometry. In aperture photometry, the $u^{*}, B, V, R_{c}, i', z', Y, J, H$, and $K_{s}$-band images are smoothed such that their point spread function (PSF) sizes are matched to the FWHM of 1.1\arcsec. We also created another set of $i'$- and $z'$-band images smoothed to match their PSFs to those of the IRAC images (FWHM $= 2$\arcsec). We measured $i' - [3.6]$, $i' - [4.5]$, $z' - [3.6]$, and $z' - [4.5]$ colors on the FWHM $= 2$\arcsec images using 2.2\arcsec diameter apertures. By adding these colors to the $i'$ or $z'$ band magnitudes measured on the FWHM $= 1.1$\arcsec images, we obtained the $3.6$ and $4.5\,\mu$m magnitudes of all detected objects. These multiband photometric measurements were corrected for Galactic extinction based on \citet{Schlegel+98} with $R_{V} = 3.1$. Our multiband photometry catalog in the SSA22-Sb1 field is the same as the ``SSA22HIT master catalog'' used in \citet{Yamanaka+20}.

\subsection{Target Selection}\label{sec:target_selection}

\begin{figure*}
\begin{center}
\includegraphics[width=1.0\linewidth, angle=0]{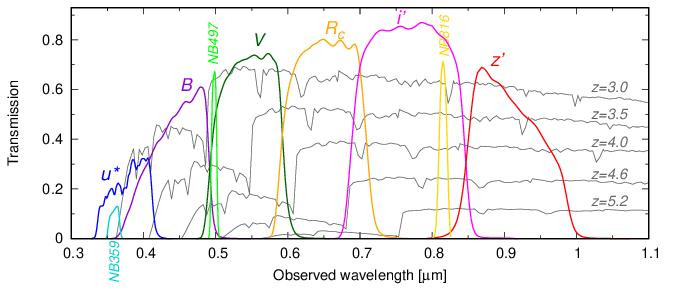}
\caption{Transmissions of the filters used in the target selection (blue: $u^*$, purple: $B$, dark-green: $V$, orange: $R_c$, magenta: $i'$, red: $z'$, cyan: NB359, light-green: NB497, and yellow: NB816) together with model spectra of star-forming galaxies at $z = 3.0, 3.5, 4.0, 4.6$, and $5.2$ from the \citet{BruzualCharlot03} library (gray lines). \label{fig:modelspec_Tsel}}
\end{center}
\end{figure*}

\begin{figure*}
\begin{center}
\includegraphics[width=1.0\linewidth, angle=0]{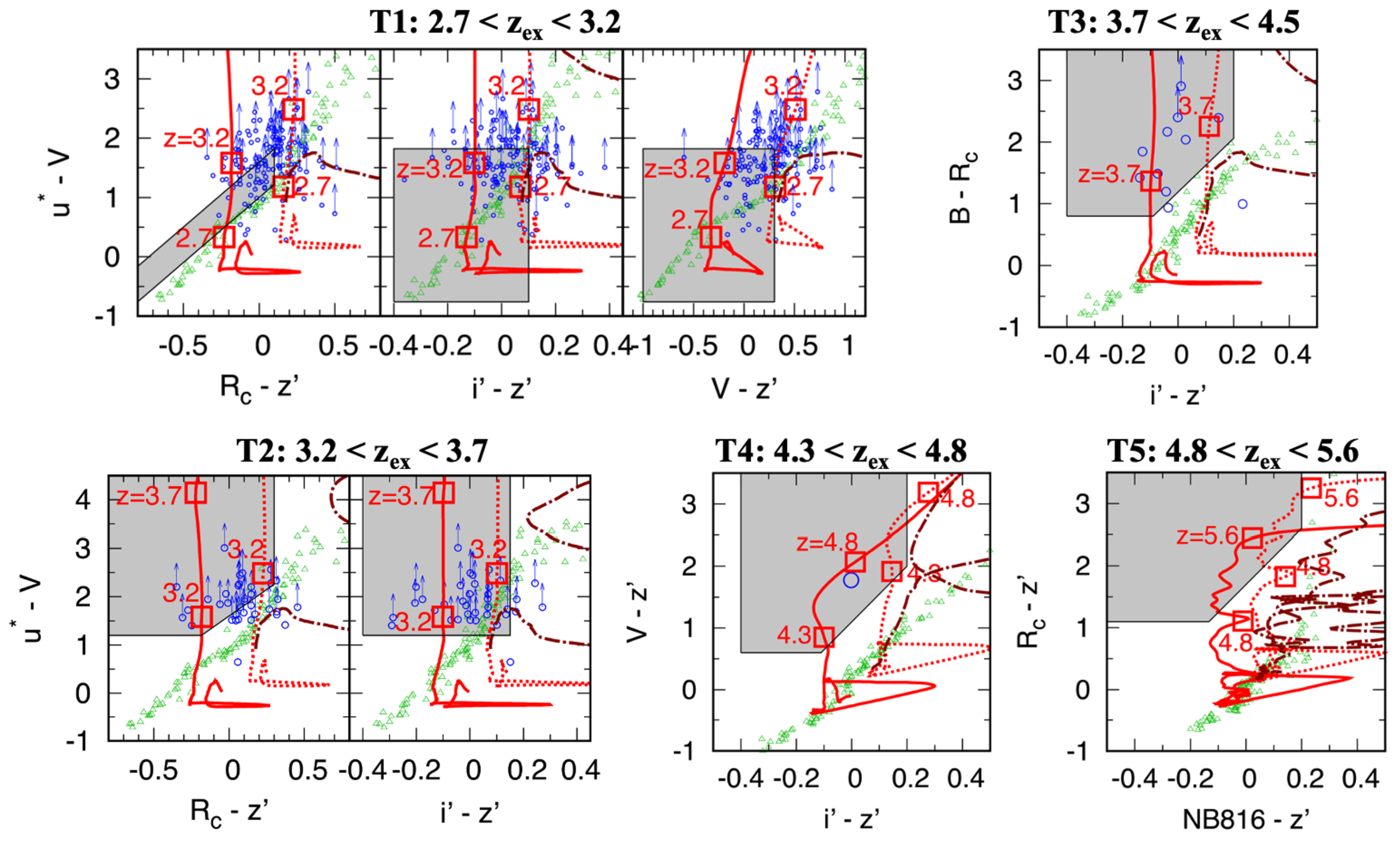}
\caption{{Two color diagrams used to select the spectroscopic targets in the five categories: T1, T2, T3, T4, and T5. In individual panels, the red solid (dotted) lines correspond to the color tracks of the \citet{BruzualCharlot03} galaxy models from $z = 0$--$7$ with constant star-formation history, stellar age of 100\,Myr, and dust extinctions of $E(B-V) = 0$ (0.25). The boundary redshifts of the categories are represented by squares. The brown dotted-dashed lines are also the \citet{BruzualCharlot03} model tracks but with instantaneous star-formation history, stellar age of 1\,Gyr, and no dust extinction. The model galaxies are dominated by the mature stellar population. The colors of Galactic stars from the \citet{Pickles98} library are shown by green triangles. The blue circles correspond to the observed galaxies whose redshifts are spectroscopically confirmed to be in the redshift range of each category \citep{Steidel+03,Kousai11,Saez+15,Hayashino+19}. The arrows indicate the $2\sigma$ limits for nondetections in the bands at wavelengths shorter than the Lyman break. Our target selection criteria are shown in the gray shaded regions. \label{fig:CC_Tsel}}}
\end{center}
\end{figure*}

We define five major categories for spectroscopic targets based on their expected redshifts {($z_{\rm ex}$)}. 
Categories T1, T2, and T3 consist of galaxies at $2.7 < z_{\rm ex} < 3.2$, $3.2 < z_{\rm ex} < 3.7$, and $3.7 < z_{\rm ex} < 4.5$, respectively, which can be used as background light sources for the H\,{\sc i} tomography at $2.5 \lesssim z \lesssim 4$. In particular, we prioritize the T2 category, which provides us with the spectra probing the HI absorption in the $z = 3.1$ protocluster. Categories T4 and T5 are defined to be at $4.3 < z_{\rm ex} < 4.8$ and $4.8 < z_{\rm ex} < 5.6$, respectively, which are used to search for galaxy overdensity regions at higher redshifts. 

We developed our own color criteria for these categories. They were more carefully designed than the commonly used dropout (e.g., \citealt{Steidel+95,Giavalisco02}) and BX \citep{Steidel+04} selection criteria such that they were more sensitive to the specific redshift intervals of our target categories. The filters used in the color selection and model galaxy spectra (\citealt{BruzualCharlot03}; hereafter, BC03) with representative redshifts of the categories are shown in Figure~\ref{fig:modelspec_Tsel}. For BC03 model templates, we assumed a Chabrier initial mass function citep{Chabrier03} with a mass range of $0.1$ -- $100\,M_{\odot}$. For dust and IGM attenuation, we applied the Calzetti law \citep{Calzetti+00} and the analytic model of \citet{Inoue+14}, respectively. Figure~\ref{fig:CC_Tsel} shows the color behaviors of the BC03 model galaxies, Galactic stars \citep{Pickles98}, and observed galaxies whose redshifts were spectroscopically confirmed by previous studies \citep{Steidel+03,Kousai11,Saez+15,Hayashino+19}. We determined the color criteria (gray shaded regions) described below to select the BC03 model galaxies with $10 <$ age [Myr] $< 100$ and $E(B-V) \lesssim 0.2$. 

For T1, we adopted 
\begin{eqnarray}
23.5 \le i' \le 25.5, \\
2.2 (R_{c} - z') + 1 \le u^{*} - V \le 2.2 (R_{c} - z') + 1.6, \\
-0.8 \le R_{c} - z' \le 0.1, \\
{-0.4 \le i' - z' \le 0.1,} \\
-1 \le V - z' \le 0.3.
\end{eqnarray}
The faint limit in the  $i'$-band magnitude was determined to collect sufficient background galaxies for H\,{\sc i} tomography with a mapping resolution of several cMpc. We also set a bright limit in the $i'$-band magnitude because many bright objects with $i' < 23.5$ showed low-$z$ contamination in our previous observations \citep{Kousai11,Hayashino+19} using the visible multiobject spectrograph (VIMOS; \citealt{LeFevre+03}) on the VLT. Strict $R_c - z'$, $i' - z'$, and $V - z'$ criteria are required to avoid contamination from the galactic stars and low-$z$ passive galaxies dominated by the matured stellar population (Figure~\ref{fig:CC_Tsel}) as well as possible artifacts such as satellite trails and stellar spikes that can make the colors extremely red or blue. Such magnitude and color constraints at wavelengths longer than the Lyman break were also adopted for the other categories.

For T2, we adopted   
\begin{eqnarray}
23.5 \le i' \le 25.5, \\
u^{*} - V \ge 1.2, \\
u^{*} - V \ge 2.2 (R_{c} - z') + 1.6, \\
-0.8 \le R_{c} - z' \le 0.3, \\
-0.4 \le i' - z' \le 0.15, \\
BV - NB497 \le 0.5 ,
\end{eqnarray}
where the $BV - NB497$ criterion was adopted to remove LAEs at $z = 3.1$. In the T1 and T2 target selections, we avoided imposing a criterion for a $B$-band depression compared to the longer-wavelength bands. For $z \sim 3$ galaxies, the $B$ band samples the wavelength range between the Ly$\alpha$ and Ly$\beta$ (Figure~\ref{fig:modelspec_Tsel}) where the foreground Ly$\alpha$ absorption is imprinted. Because the main aim of this work is to collect the background sight-lines for H\,{\sc i} absorption analysis, we avoided introducing any bias in the $B$-band flux due to the foreground absorption. 

For T3, we adopted 
\begin{eqnarray}
24 \le i' \le 25.5,\\
B - R_{c} \ge 0.8,\\ 
B - R_{c} \ge 4.3 (i' - z') + 1.2,\\
-0. 4 \le i' - z' \le 0.2, \\
u^{*} > 26.9 (2\sigma), \\
NB359 > 26.6 (2.5\sigma). 
\end{eqnarray}

For T4, we adopted 
\begin{eqnarray}
24 \le z'  \le 25.8, \\
V - z' \ge 0.6, \\
V - z' \ge 4.5(i' - z') + 1.1, \\
-0.4 \le i' - z' \le 0.2, \\ 
B > 28.35 (1.5\sigma).
\end{eqnarray}

For T5, we adopted 
\begin{eqnarray}
z' \le 25.75, \\
R_{c} - z' \ge 1.1, \\
R_{c} - z' \ge 4.5 (NB816 - z') + 1.7, \\
NB816 - z' \le 0.2, \\
B > 28.35 (1.5 \sigma).
\end{eqnarray}
Note that the above criteria were slightly relaxed for a small number of objects so that we could effectively fill the spectroscopic slits. We made it possible to distinguish the objects selected by the relaxed color criteria in our published catalog (see Appendix~\ref{sec:ap_HITcat}). 

Galaxies that are brighter than $25.5$\,mag in the $i'$ band and spectroscopically confirmed as $z_{\rm spec} > 3.2$ in previous studies (\citealt{Steidel+03,Kousai11,Saez+15,Hayashino+19}; see also Section~\ref{sec:otherdata}) were added to the T2 and T3 categories, which are referred to as T2prevz and T3prevz, respectively. Unless specified otherwise, the T2/T3 categories include the T2prevz/T3prevz subcategories throughout this paper. We also estimated the photometric redshifts (photo-$z$) for the $i'$- or $z'$-band detected objects using a public software ``HYPERZ'' \citep{Bolzonella+00} to increase spectroscopic target candidates. In this photo-$z$ estimation for the target selection, we did not use the three Subaru narrowband data. We used composite spectral templates of stellar components (BC03) and nebular emission lines \citep{Inoue+11b}, which are the same as those used in \citet{Mawatari+16b}. 

\begin{figure*}[]
\begin{center}
\includegraphics[width=1.0\linewidth, angle=0]{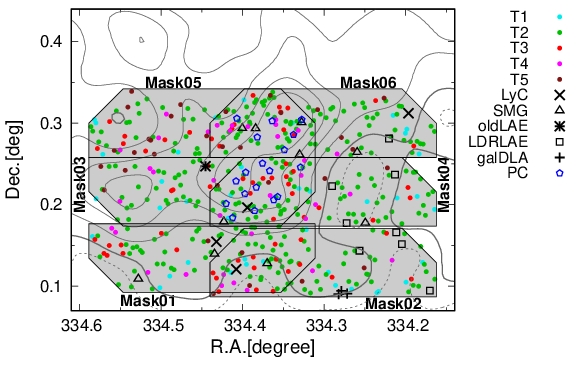}
\caption{Sky coverage of the six DEIMOS masks (gray shade) in the SSA22-Sb1 field. All spectroscopic targets for the SSA22-HIT survey are superposed by their categories: T1 ($2.7 < z_{\rm ex} < 3.2$) in cyan filled circles, T2 ($3.2 < z_{\rm ex} < 3.7$) in green filled circles, T3 ($3.7 < z_{\rm ex} < 4.5$) in red filled circles, T4 ($4.3 < z_{\rm ex} < 4.8$) in magenta filled circles, T5 ($4.8 < z_{\rm ex} < 5.6$) in brown filled circles, LyCs by x-marks, SMGs by open triangles, oldLAE by an asterisk, LAEs in the low-density region (LDRLAEs) by open squares, $z = 3.1$ protocluster members (PCs) by open pentagons, and the galaxy sight-line DLA and its counterpart candidates (galDLAs) by crosses. The background contours show the smoothed number density of the $z = 3.1$ LAEs: 0.5 (dashed), 1 (thick), 2, 3, 4 and 5 times the average density (0.2\,arcmin$^{-2}$; \citealt{Yamada+12a}).  \label{fig:skydist_targets}}
\end{center}
\end{figure*}

We used the DSIMULATOR\footnote{\url{https://www2.keck.hawaii.edu/inst/deimos/dsim.html}} software to design six DEIMOS slit masks (Mask01, 02, 03, 04, 05, and 06). The six masks were arranged within $26 \times 15$\,arcmin$^2$ of the SSA22-Sb1 field, and they overlap in the central $7.7 \times 15$\,arcmin$^2$ (Figure~\ref{fig:skydist_targets}). We achieved a doubled surface density of the targets in the overlapped area where no target was shared among the different masks. We prioritized the categories as T2 and T3 with previously confirmed redshift (T2prevz and T3prevz) $>$ T5 $>$ T2 $>$ T4 $>$ T3 $>$ T1, considering their number densities. We also prioritized the targets in each category in the following order: objects satisfying both the color criteria and the photo-$z$ range, objects satisfying the color criteria, and objects satisfying only the photo-$z$ range. 
Eventually, on average, $85$ slits per mask were assigned, excluding alignment stars. Out of the total 513 slits, $8$\,\%, $58$\,\%, $12$\,\%, $7$\,\%, and $6$\,\% are in T1, T2, T3, T4, and T5, respectively. 

A small part of the slits were also assigned to the following interesting targets: candidate Lyman continuum leakers identified with NB359 (LyC; \citealt{Iwata+09}), SMGs detected with AzTEC and Atacama Large Milimeter/submillimeter Array (ALMA; \citealt{Umehata+14,Umehata+18}), LAEs with the $K_s$-band detection indicating old stellar age (oldLAEs; \citealt{Yamada+12a}), LAEs lying within the LAEs’ low-density region (LDRLAEs; \citealt{Yamada+12a}), a damped Ly$\alpha$ system identified along a galaxy sight-line (galDLA; \citealt{Mawatari+16a}), and its counterpart candidates. We also assigned the remaining slits to protocluster members (PCs) that were brighter than $25.5$\,mag in the $i'$ band and were spectroscopically confirmed as $3.04 \leq z \leq 3.12$ in previous observations (\citealt{Steidel+03,Kousai11,Saez+15,Hayashino+19}; see also Section~\ref{sec:otherdata}). The PC slits are limited to the $8 \times 10$\,arcmin$^2$ area around the $z = 3.1$ LAE density peak, which roughly corresponds to the overlapped region by Mask03, Mask04, Mask05, and Mask06. We assigned these PC slits to perform high spatial resolution H\,{\sc i} tomography at {$2.5 \lesssim z \lesssim 3$}. The sky distributions of all of the targets for the spectroscopic observations are shown in Figure~\ref{fig:skydist_targets}.

\subsection{Spectroscopic Observations}\label{sec:DEIMOS_obs}

\begin{table*}[]
\begin{center}

\caption{Summary of DEIMOS Observations in the SSA22 field} \label{tb:DEIMOSobs}
\begin{tabular}{cccccccc}
\hline
\hline
 & R.A.\tablenotemark{a} (J2000) & Decl.\tablenotemark{a} (J2000) & $N_{\rm data}$\tablenotemark{b}  & Dates\tablenotemark{c} (UT) & Exposure Time (s) & Seeing\tablenotemark{d} (arcsec) \\
\hline
Mask01 & $22^{\rm h}17^{\rm m}48.0^{\rm s}$ & $+00{\degr}08{\arcmin}04.8{\arcsec}$& 92 & 2015 Oct 12 & 7200 & $0.9$ -- $1.0$  \\
\hline
Mask02 & $22^{\rm h}17^{\rm m}12.3^{\rm s}$ & $+00{\degr}07{\arcmin}43.2{\arcsec}$ & 84 & 2015 Oct 12 & 7200 & $0.9$  \\
\hline
Mask03 & $22^{\rm h}17^{\rm m}48.0^{\rm s}$ & $+00{\degr}12{\arcmin}57.7{\arcsec}$ & 89 & 2015 Oct 10 & 7200 & $0.7$ -- $0.8$ \\
 & & & & 2016 Aug 3 -- 4 & 19,230 & $0.7$ -- $1.3$  \\
 & & & & 2016 Oct 9 & 1800 & $0.9$  \\
\hline
Mask04 & $22^{\rm h}17^{\rm m}12.3^{\rm s}$ & $+00{\degr}12{\arcmin}56.3{\arcsec}$ & 92 & 2015 Oct 10 & 7200 & $0.6$ -- $0.8$ \\
 & & & & 2016 Aug 1 -- 3 & 21,600 & $0.7$ -- $1.3$ \\
\hline
Mask05 & $22^{\rm h}17^{\rm m}48.0^{\rm s}$ & $+00{\degr}17{\arcmin}59.4{\arcsec}$ & 99 & 2015 Aug 16 -- 17\tablenotemark{e} & 3600 & $0.7$ -- $0.8$ \\
 & & & & 2015 Sep 7 & 6579 & $0.7$ -- $0.9$  \\
 & & & & 2015 Oct 13 & 3000 & $0.7$ -- $1.1$  \\
 & & & & 2016 Jul 31 -- Aug 1 & 15,600 & $1.0$ -- $1.5$  \\
\hline
Mask06 & $22^{\rm h}17^{\rm m}12.3^{\rm s}$ & $+00{\degr}17{\arcmin}59.2{\arcsec}$ & 91 & 2015 Aug 15 -- 16\tablenotemark{e} & 6868 & $0.7$ -- $1.2$ \\
 & & & & 2015 Oct 10 -- 13 & 9700 & $0.7$ -- $1.1$  \\
 & & & & 2016 Jul 31 & 12,600 & $0.9$ -- $1.1$ \\
 & & & & 2016 Oct 8 -- 9 & 19,500 & $0.8$ -- $1.0$  \\
\hline
\end{tabular}
\end{center}
\tablenotetext{a}{The mask R.A. and decl. coordinates are defined as the geometrical center positions of the 16.7\arcmin $\times$ 5\arcmin fields-of-view.}
\tablenotetext{b}{The number of slits for which we actually obtained spectral data. Slits that are designed but fall into the CCD gap are not included in $N_{\rm data}$, while slits for the alignment stars are included.}
\tablenotetext{c}{Wavelength and flux calibrations were performed for each dataset of the continuous observing dates. }
\tablenotetext{d}{Seeing sizes were measured for raw spectra of the alignment stars at $\lambda \sim 5500$\,\AA. }
\tablenotetext{e}{The background sky level was very high due to the cloudy conditions in 2015 Aug.}
\end{table*}

We performed  spectroscopic observations of the targets described above with DEIMOS \citep{Faber+03} on the Keck II telescope in 2015 and 2016 using S274D (PI: T. Yamada), U066D (PI: D. Schlegel), S313D (PI: T. Yamada), and S290D (PI: T. Yamada). We were awarded four full nights and 10 half nights in total. Because some nights were lost due to bad weather, the effective observation time corresponded to $41$\, h. 

We used a 600ZD grating with a slit width of $1$\arcsec, which yielded a spectral resolution of $\Delta \lambda \sim 4.7$\,\AA\ or resolving power $R \sim 1000$. The pixel scales along the wavelength and spatial directions are $\sim 0.62$\,\AA\,pix$^{-1}$ and $0.1185$\arcsec\,pix$^{-1}$, respectively. 
We chose the GG400 order blocking filter and set a central wavelength of $6300$\,\AA\ or $6500$\,\AA, which provided a wavelength coverage of $4000$\,\AA\ $\lesssim \lambda \lesssim 9000$\,\AA. The observed coordinates, number of slits, date, exposure time, and seeing are listed in Table~\ref{tb:DEIMOSobs}. The raw data were classified by their slit masks and continuous observation dates (observing dates (observing ``runs''). The data reduction of each slit mask and each observing run was performed separately (see Section~\ref{sec:datareduc} and Table~\ref{tb:DEIMOSobs}). 

The DEIMOS standard procedure for flat and arc data is not optimal for our main science focusing on the H\,{\sc i} Ly$\alpha$ absorptions at $\lambda \lesssim 5500$\,\AA\ because of the low counts and limited number of arc lines at the blue wavelengths. Instead, we used two types of flat and arc data optimized for the red and blue wavelengths using the operational script ``calib\_blue''\footnote{\url{https://www2.keck.hawaii.edu/inst/deimos/procs/calib_blue.html}}. Two types of spectroscopic standard stars, blue early-type stars (Feige110, BD$+$33d2642, BD$+$25d4655, and Hz44) and red late-type stars (Hilt102 and Cyg OB2 No.9.) were observed to correct for contamination from second-order light in the flux calibration (see Section~\ref{sec:datareduc}).

\section{Data Reduction}\label{sec:datareduc}

For data reduction, we ran the commonly used DEEP2 DEIMOS spec2d pipeline \citep{Cooper+12,Newman+13} version 1.1.4 for every observing run data. In the flat-fielding and wavelength calibration, red/blue flat and arc data were adopted for the red/blue CCD chips separately. Using sky lines, we further corrected the systematic wavelength offsets after calibration with the arc data. 
We eventually achieved an rms of the wavelength calibration of less than $0.3$\,\AA\ over the entire wavelength coverage. 
After stacking all frames in each observing run and each mask, the pipeline subtracts the background sky and extracts one-dimensional spectra from two-dimensional spectral images. We chose the spatial summation widths of $\sim$ seeing-FWHM in the individual observation runs to maximize the signal-to-noise ratio (S/N) for the continuum. The flux loss due to the slit and narrow spatial summation width was not corrected. 

Flux calibration was performed with IRAF. We found that second-order light at $3800$\,\AA\ $\lesssim \lambda \lesssim 4500$\,\AA\ contaminates the main first-order spectra at $\lambda \ga 8000$\,\AA. We performed flux sensitivity functions at $\lambda \le 7500$\,\AA\ and $\lambda \ge 7500$\,\AA\ separately using blue and red standard stars. The intrinsic flux of the red standard stars at $\lambda \lesssim 4500$\,\AA\ makes the second-order light contamination negligible, while it is too faint to obtain a reliable sensitivity function at blue wavelengths. We combined the sensitivity functions made from the blue and red standard stars and used them for flux calibration. After the flux calibration, we stacked all observed runs' spectra for each mask. The $2\sigma$ depths at $\lambda = 5000$\,\AA\ reached $0.19$ -- $0.48$\,$\mu$Jy per resolution (first column in Table~\ref{tb:HITzsum}). We note that the noise $\sigma$ increases with decreasing wavelengths at $\lambda < 5000$\,\AA. The noise level increases by a factor of $\sim 2$ from $\lambda = 4750$--$4250$\,\AA, which is in contrast to the constant noise level at $5000$\,\AA\ $< \lambda < 7000$\,\AA. The spectra were corrected for the Galactic dust extinction based on \citet{Schlegel+98} with $R_V = 3.1$.

The spectra reduced with the spec2d pipeline are provided on an air wavelength scale. We converted the wavelengths from air to vacuum following a formula of \citet{Greisen+06}, which is the same as that adopted by the International Union of Geodesy and Geophysics General Assembly (1999)\footnote{\url{http://www.iugg.org/assemblies/1999birmingham/1999crendus.pdf}}.

\section{Redshift Catalog Construction}\label{sec:redshift}

\subsection{Redshift Determination in SSA22-HIT}\label{sec:zid_HIT}

\begin{figure}
\begin{center}
\includegraphics[width=1.0\linewidth, angle=0]{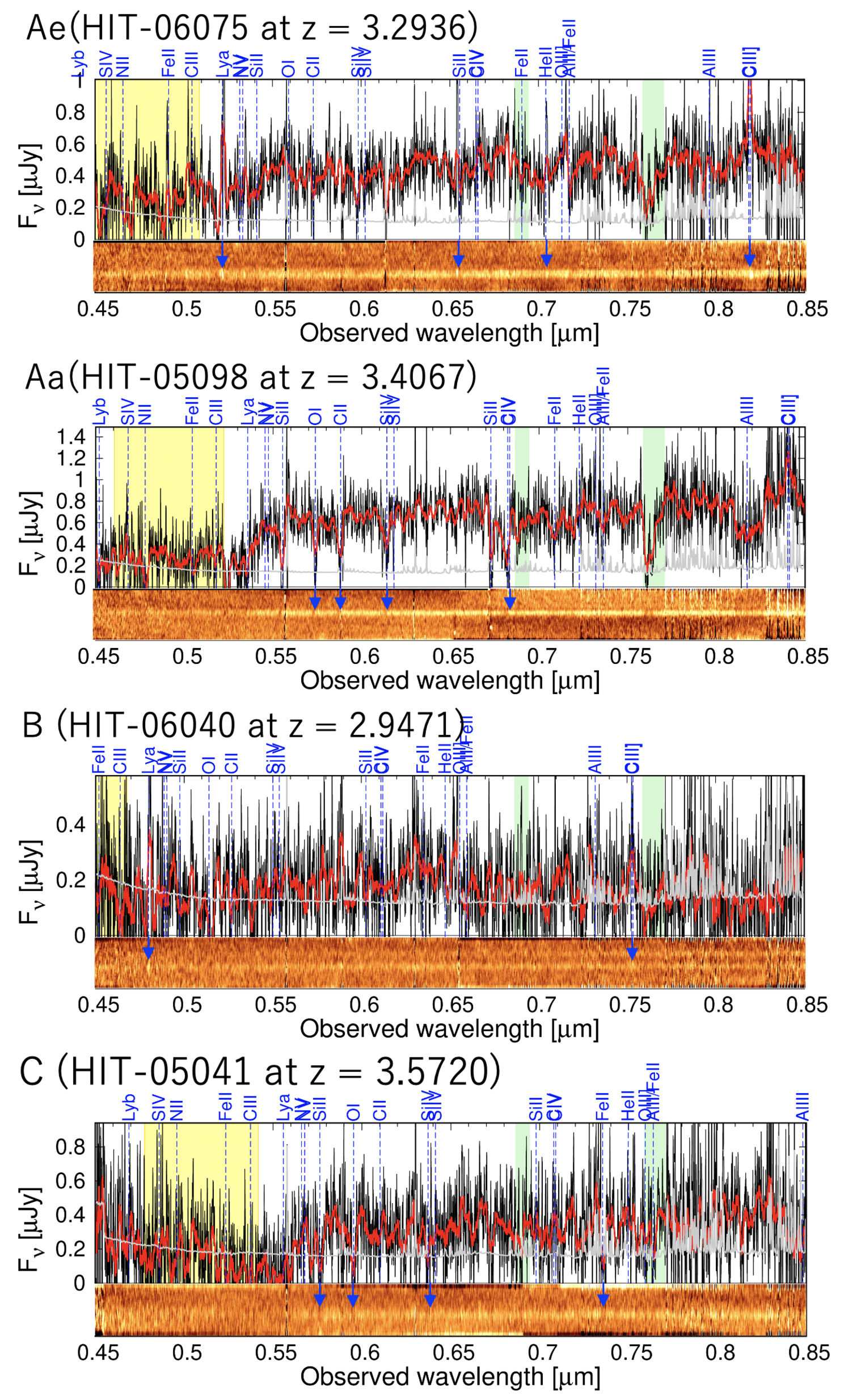}
\caption{Observed spectra of four example galaxies taken with the Keck/DEIMOS in the SSA22-HIT survey. The four galaxies are sorted by their redshift quality flags. In each figure, the one-dimensional (top panel) and two-dimensional (bottom panel) spectra are shown. The black and gray lines correspond to the one-dimensional spectra and error spectra that are smoothed with a 5 pixel box-car kernel, respectively. We also show much more smoothed spectra with a 39 pixel box-car kernel using the red lines. The yellow shaded region corresponds to the wavelength range in which we analyze the foreground H\,{\sc i} Ly$\alpha$ absorption. The green shaded regions represent the atmospheric absorption bands. The vertical dashed lines represent the wavelengths of the possible emission/absorption lines. The blue arrows superimposed on the two-dimensional spectra mark the lines that were visually identified. \label{fig:HITspec_example}}
\end{center}
\end{figure}

\begin{figure}
\begin{center}
\includegraphics[width=1.0\linewidth, angle=0]{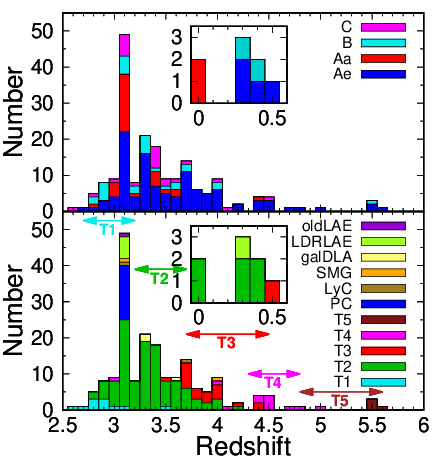}
\caption{Spectroscopic redshift distribution of the objects identified in the SSA22-HIT survey by the redshift quality flags (top) and the target categories (bottom). The objects that show low-$z$ contamination at $0 \leq z \leq 0.5$ are shown in the subpanels. The two-headed arrows represent the expected redshift ranges of the five major target categories (T1, T2, T3, T4, and T5). \label{fig:zdist_HIT}}
\end{center}
\end{figure}

\begin{figure}
\begin{center}
\includegraphics[width=1.0\linewidth, angle=0]{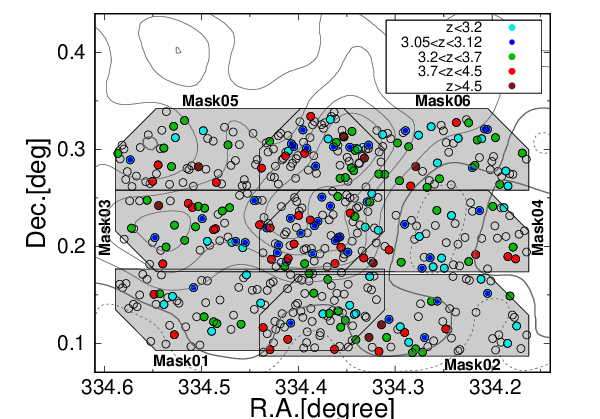}
\caption{Sky distributions of the objects observed in the SSA22-HIT survey. All targets observed are shown by open circles, while the filled circles correspond to the objects whose redshifts are determined ($z < 3.2$ in cyan, $3.05 \leq z \leq 3.12$ in blue, $3.2 \leq z < 3.7$ in green, $3.7 \leq z < 4.5$ in red, and $z \geq 4.5$ in brown). The background contours show the $z = 3.1$ LAE number density, which are the same as those in Figure~\ref{fig:skydist_targets}.  \label{fig:skydist_HIT}}
\end{center}
\end{figure}

\begin{table*}[]
\begin{center}
\caption{Summary of Redshift Determination in the SSA22-HIT Survey for Each Slit Mask} \label{tb:HITzsum}
\begin{tabular}{cccccccc}
\hline
\hline
 Mask ID & Depth\tablenotemark{a} & $N_{\rm sci}$\tablenotemark{b} & $N_{\rm Ae}$\tablenotemark{c} & $N_{\rm Aa}$\tablenotemark{c} & $N_{\rm B}$\tablenotemark{c} & $N_{\rm C}$\tablenotemark{c} & {Success Rate\tablenotemark{d}} \\
  & ($\mu$Jy) &  &  &  &  & & (\%) \\
\hline
mask01 & 0.48 & 86 & 11 & 2 & 5 & 3 & 24.4 \\
mask02 & 0.40 & 78 & 20 & 4 & 5 & 1 & 38.5 \\
mask03 & 0.22 & 84 & 22 & 5 & 5 & 6 & 45.2 \\
mask04 & 0.22 & 86 & 25 & 9 & 4 & 3 & 47.7 \\
mask05 & 0.25 & 93 & 17 & 6 & 2 & 4 & 31.2 \\
mask06 & 0.19 & 86 & 18 & 6 & 9 & 6 & 45.3 \\
\hline
Total & --- & 513 & 113 & 32 & 30 & 23 & 38.6 \\
\hline
\end{tabular}
\end{center}
\tablenotetext{a}{The $2\,\sigma$ depth per resolution ($\Delta \lambda \sim 4.7$\,\AA) at $\lambda = 5000$\,\AA.}
\tablenotetext{b}{Number of slits for the science targets except for the alignment stars.}
\tablenotetext{c}{Number of objects with a given redshift quality flag (see the text).}
\tablenotetext{d}{{Success rate defined as the number fraction of objects whose redshifts are confirmed with the quality flags of Ae, Aa, B, or C over the science targets: $= (N_{\rm Ae}+N_{\rm Aa}+N_{\rm B}+N_{\rm C}) / N_{\rm sci}$.}}
\end{table*}

\begin{figure}
\begin{center}
\includegraphics[width=1.0\linewidth, angle=0]{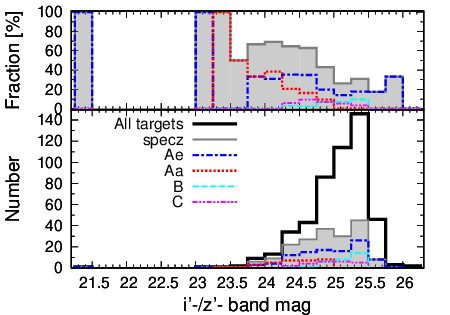}
\caption{{(Bottom) Magnitude distribution of all targets (black solid) and objects spectroscopically confirmed (grey shaded) in the SSA22-HIT survey. We used the $i'$ band magnitudes. Because some of the T4 and T5 category targets and SMGs were not detected in the $i'$-band, we used the $z'$-band magnitudes for them. The spectroscopically confirmed objects were divided by their redshift quality flags. (Top) Number fraction of spectroscopically confirmed objects over all targets. } \label{fig:maghist_HIT}}
\end{center}
\end{figure}

\begin{figure*}
\begin{center}
\includegraphics[width=1.0\linewidth, angle=0]{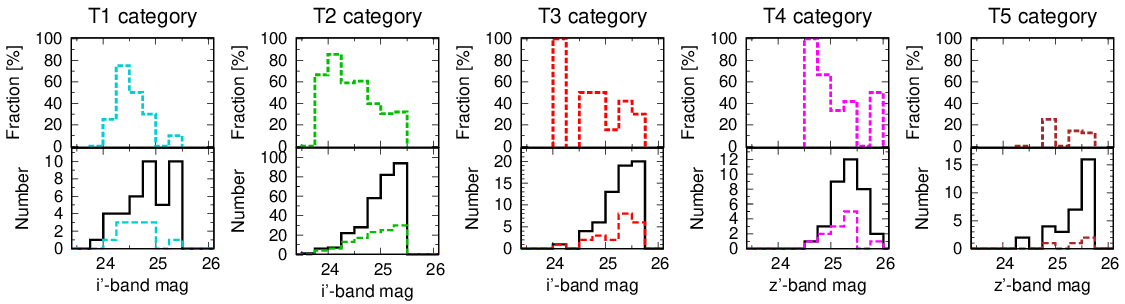}
\caption{{Magnitude distribution (bottom) and number fraction of the spectroscopically confirmed objects (bottom) for each of the five major categories. In each panel, the solid and dashed histograms represent the observed and spectroscopically confirmed objects in the given category, respectively. We used the $i'$-band magnitudes for the T1, T2, and T3 categories, while the $z'$-band magnitudes are used for the T4 and T5 categories.} \label{fig:maghist_HITcate}}
\end{center}
\end{figure*}

\begin{table*}[]
\begin{center}
\caption{Summary of Redshift Determination in the SSA22-HIT Survey for Each Category} \label{tb:HITzsum_T}
\begin{tabular}{ccccccccc}
\hline
\hline
 Category & $N_{\rm sci}$\tablenotemark{a} & $N_{\rm Ae}$\tablenotemark{a} & $N_{\rm Aa}$\tablenotemark{a} & $N_{\rm B}$\tablenotemark{a} & $N_{\rm C}$\tablenotemark{a} & {Success Rate\tablenotemark{a}} & $N_{\rm zcorrect}$\tablenotemark{b} & Recovery Rate\tablenotemark{c} \\
  &  &  &  &  &  & (\%) &  & (\%) \\
\hline
T1 & 40 & 6 & 1 & 2 & 2 & 27.5 & 9 & 81.8 \\
T2\tablenotemark{d} & 270 & 39 & 16 & 26 & 15 & 35.6 & 38 & 39.6 \\
T2prevz\tablenotemark{d} & 29 & 14 & 7 & 1 & 1 & 79.3 & --- & --- \\
T3\tablenotemark{d} & 56 & 15 & 0 & 0 & 1 & 28.6 & 12 & 75.0 \\
T3prevz\tablenotemark{d} & 7 & 6 & 0 & 0 & 0 & 85.7 & --- & --- \\
T4 & 35 & 8 & 1 & 0 & 3 & 34.3 & 8 & 66.7 \\
T5 & 32 & 3 & 0 & 1 & 0 & 12.5 & 3 & 75.0 \\
LyC & 4 & 2 & 0 & 0 & 0 & 50.0 & --- & --- \\
SMG & 10 & 1 & 0 & 0 & 1 & 20.0 & --- & --- \\
oldLAE & 1 & 1 & 0 & 0 & 0 & 100.0 & --- & --- \\
LDRLAE & 8 & 7 & 0 & 0 & 0 & 87.5 & --- & --- \\
galDLA & 3 & 3 & 0 & 0 & 0 & 100.0 & --- & --- \\
PC & 18 & 8 & 7 & 0 & 0 & 83.3 & --- & --- \\
\hline
\end{tabular}
\end{center}
\tablenotetext{a}{The definitions of $N_{\rm sci}$, $N_{\rm Ae}$, $N_{\rm Aa}$, $N_{\rm B}$, $N_{\rm C}$, and the success rate are the same as those in Table~\ref{tb:HITzsum}. }
\tablenotetext{b}{Number of each category of objects whose spectroscopic redshifts are found to be within the expected range.}
\tablenotetext{c}{Recovery rate defined as a number fraction of objects whose redshifts are within the expected range of each category over those spectroscopically confirmed redshifts (irrespective of the expected range): $= N_{\rm zcorrect} / (N_{\rm Ae}+N_{\rm Aa}+N_{\rm B}+N_{\rm C})$}
\tablenotetext{d}{In this table, the T2/T3 category objects are divided into those with and without previous spectroscopic redshift measurements (see also Section~\ref{sec:target_selection}) that are referred to as T2prevz/T3prevz and just T2/T3, respectively. }
\end{table*}

We determined the spectroscopic redshifts of the target objects using a ``specpro'' software \citep{Masters+11}, which adopts cross-correlation routines for the observed and template spectra. The observed spectra were smoothed with a $5$ pixel box-car kernel for cross-correlation. We used the following 10 spectral templates: 
\begin{itemize}
\item VLBG: an LBG template from VIMOS VLT Deep Survey (VVDS; \citealt{LeFevre+05}).
\item SLBG: a composite spectrum of $z \sim 3$ LBGs \citep{Shapley+03}. 
\item VI08eLBG: a composite spectrum of LBGs with strong Ly$\alpha$ emission line from the previous VIMOS survey \citep{Hayashino+19}. 
\item VI08aLBG: a composite spectrum of LBGs with no Ly$\alpha$ emission line from the previous VIMOS survey \citep{Hayashino+19}. 
\item HITdr1eLBG: a composite of the SSA22-HIT spectra of LBGs with strong Ly$\alpha$ emission line, of which redshifts are easily identified (e.g., they are already confirmed in the previous works; multiple emission/absorption lines are obviously seen). 
\item HITdr1aLBG: a composite of the SSA22-HIT spectra of LBGs with no Ly$\alpha$ emission line, of which redshifts are easily identified (e.g., they are already confirmed in the previous works; multiple emission/absorption lines are obviously seen). 
\item VSBurst: a low-$z$ starburst galaxy template with extremely strong nebular lines from the VVDS.
\item VEs: a low-$z$ elliptical galaxy template from the VVDS.
\item SDSSQSO: a broad line quasar template from the Sloan Digital Sky Survey \citep{Schneider+10}. 
\item M6star: an M6 type stellar template \citep{Pickles98}.
\end{itemize} 
The VLBG, SLBG, VSBurst, VEs, SDSSQSO, and M6star templates are the same as those in the original specpro library \citep{Masters+11}.

For many observed galaxies at $z > 2$, similar redshift solutions can be obtained by multiple templates among VLBG, SLBG, VI08eLBG, VI08aLBG, HITdr1eLBG, and HITdr1aLBG. To determine the best-fit template, we set the priority as follows. The SLBG, HITdr1eLBG, and HITdr1aLBG templates are more prioritized than the VLBG, VI08eLBG, and VI08aLBG templates because the former have spectral resolutions comparable to the observed DEIMOS spectra. The HITdr1eLBG and HITdr1aLBG templates, which are generated from subsets of the observed objects to be fitted, are less prioritized than the SLBG template. As a result, most of the spectroscopically confirmed galaxies at $z > 2$ with and without the Ly$\alpha$ emission line were fit by the SLBG and HITdr1aLBG templates, respectively. 

We also visually inspected the spectra because the cross-correlation scheme sometimes fails to find the redshift solution, as suggested by the code developers \citep{Masters+11}. The spectral features that we carefully searched for are the Ly$\alpha$ emission/absorption ($1215.67$\,\AA\ in the rest frame\footnote{{The rest-frame wavelengths of the emission/absorption lines are extracted from \citet{NIST_ASD} in a vacuum below 2000\,\AA\ or otherwise in air. Note that these values are for reference only and are not used to calculate the redshifts.}}) line, He\,{\sc ii} ($1640.35$\,\AA), and C\,{\sc iii}] ($1906.68$\,\AA\ and $1908.73$\,\AA) emission lines, and Si\,{\sc ii} ($1260.42$\,\AA), O\,{\sc i} ($1302.17$\,\AA), Si\,{\sc ii} ($1304.37$\,\AA), C\,{\sc ii} ($1334.53$\,\AA), Si\,{\sc iv} ($1393.76$\,\AA\ and $1402.77$\,\AA), Si\,{\sc ii} ($1526.71$\,\AA), C\,{\sc iv} ($1548.19$\,\AA\ and $1550.77$\,\AA), Fe\,{\sc ii} ($1608.45$\,\AA), and Al\,{\sc ii} ($1670.79$\,\AA) absorption lines. We also searched [O\,{\sc ii}] ($3727.3$\,\AA\ and $3729.2$\,\AA), H$\beta$ ($4861.3$\,\AA), [O\,{\sc iii}] ($4958.9$ and $5006.8$\,\AA), and H$\alpha$ ($6562.8$\,\AA) emission lines as a signature of low-redshift contamination. In the visual inspection, we set four types of flags for the determined redshifts based on their reliability and their spectral features (redshift quality flags): Ae (firm redshifts are obtained from clear emission lines), Aa (firm redshifts are obtained from clear absorption lines), B (probable redshifts are obtained from emission lines), and C (possible redshifts are obtained from absorption lines). The flag C objects should be treated carefully for scientific use, because they may include redshift misidentifications. Figure~\ref{fig:HITspec_example} shows the example spectra of the four redshift-quality flags. All of the spectra of the spectroscopically confirmed objects are shown in Appendix~\ref{sec:ap_HITspec}.

We eventually obtained spectroscopic redshifts for 198 objects, of which the redshift and sky distributions are shown in Figures~\ref{fig:zdist_HIT} and \ref{fig:skydist_HIT}, respectively. Table~\ref{tb:HITzsum} summarizes their detailed numbers using masks and redshift quality flags. Success rates in the redshift determination, which are defined as the number fractions of objects whose redshifts are confirmed over all of the observed targets, are roughly inversely proportional to the depths. Figure~\ref{fig:maghist_HIT} shows the broadband magnitude histograms of spectroscopically confirmed objects as well as for all targets. The fainter the continuum of an object, the harder it is to determine its redshift. This tendency is more pronounced in objects identified by absorption lines (Aa and C) than in those identified by emission lines (Ae and B). Figure~\ref{fig:maghist_HITcate} also shows the magnitude histograms for each of the five major categories. Their magnitude dependencies in the success rates seem similar to those of all objects, which is, however, not statistically significant except for the T2 category. The number summary for each target category is listed in Table~\ref{tb:HITzsum_T}. To evaluate the accuracy of the prior redshift expectations ($z_{\rm ex}$) for the targets in the five major categories (T1, T2, T3, T4, and T5; Section~\ref{sec:target_selection}), we define a recovery rate as the number fraction of each category of objects whose spectroscopic redshifts are found to be within the expected range over those spectroscopically confirmed redshifts (irrespective of the expected redshift range). To calculate the recovery rates, we excluded the objects with previous spectroscopic redshift measurements in the T2/T3 categories (T2prevz/T3prevz; Section~\ref{sec:target_selection}). The recovery rates in T1, T2, T3, T4, and T5 were 82\,\%, 40\,\%, 75\,\%, 67\,\%, and 75\,\%, respectively. As shown in Figure~\ref{fig:zdist_HIT}, the spectroscopic redshift distribution in each major category extends beyond the low and high boundaries of the prior expectation. 
We have published the properties of all of the observed targets, including the redshift measurements, redshift quality flags, and target categories (see Appendix~\ref{sec:ap_HITcat}). 

The cross-correlation technique is expected to provide systemic redshifts unless the spectral properties of the templates used are very different from those of observed objects. The majority of our redshift measurements are determined by cross-correlation with the SLBG and HITdr1aLBG templates. The SLBG template is a composite of LBGs whose redshift and magnitude ranges are similar to those of our HIT sample \citep{Shapley+03,Steidel+03}. We then consider our redshift measurements to be systemic. 

The error estimation of the redshifts measured using the cross-correlation technique is not trivial, a conservative estimate for the redshift uncertainty can be given by the spectral resolution of our HIT data ($\Delta \lambda \sim 4.7$\,\AA; see Section~\ref{sec:DEIMOS_obs}), resulting in $\Delta z \sim 0.001(1+z)$ or $\Delta v \sim 280$\,km\,s$^{-1}$. After comparing with previous redshift measurements from the literature and archival catalogs (see Sections~\ref{sec:otherdata}), there were 148 objects whose redshifts were newly determined in our SSA22-HIT survey. For the remaining 50 objects, we investigated the redshift differences between those measured in the present study and those of previous works, $z_{\rm HIT} - z_{\rm previous}$, where the latter redshifts are also observed or expected systemic ones (Sections~\ref{sec:otherdata}). Some objects were observed in multiple previous studies, and there were 84 pairs of $z_{\rm HIT}$ and $z_{\rm previous}$. Their redshift differences were found to be $\approx 0.0002$ on average, with a standard deviation of $\approx 0.004$. Considering the redshift uncertainties of our and previous measurements (see Section~\ref{sec:otherdata} and Table~\ref{tb:speczcomp}), we conclude that our redshift measurements are consistent with those of previous studies.

\subsection{Ancillary Data}\label{sec:otherdata}

\begin{table*}[]
\begin{center}
\caption{Summary of Redshift Catalogs Available in the SSA22-Sb1 Field} \label{tb:speczcomp}
\begin{tabular}{cccccccc}
\hline
\hline
Catalog ID & Instrument & Wavelength & Resolution & {$z$ Unc.} & $N$\tablenotemark{c} & Object Type & References\tablenotemark{d} \\
 &  & Coverage & $\Delta \lambda$\tablenotemark{a}  & {$\Delta z$\tablenotemark{b}} & & & \\
\hline
SSA22-HIT & Keck/DEIMOS & $4000$ -- $9000$\,\AA\ & 4.7\,\AA\ & {0.004} & 175\tablenotemark{e} & LBG,LyC,SMG,LAE & This work \\
VIMOS06  & VLT/VIMOS & $3700$ -- $6800$\,\AA\ & 25\,\AA\ & {0.02} & 50\tablenotemark{e} & LBG,AGN & (1),This work \\
VIMOS08  & VLT/VIMOS & $3700$ -- $6800$\,\AA\ & 25\,\AA\ & {0.006} & {82} & {LBG,AGN} & (1),(2) \\
VIMOS12  & VLT/VIMOS & $3700$ -- $6800$\,\AA\ & 25\,\AA\ & {0.02} & 39 & LBG,SMG,LyC & (3),This work \\
DEIMOS08  & Keck/DEIMOS & $4000$ -- $8000$\,\AA\ & 2.8\,\AA\ & {0.002} & 34\tablenotemark{e} & LAE,LBG,LAA,Kdet & This work \\
oIMACS  & Magellan/IMACS & $4800$ -- $7800$\,\AA\ & $6.7$\,\AA\ & {0.006} & 29 & LAE,LyC,LAB,LBG & This work \\
Steidel+03  & Keck/LRIS & $4000$ -- $7000$\,\AA\ & $7.5$\,\AA\ & {0.0035} & 133 & {LBG,AGN} & (4) \\
Nestor+13  & Keck/LRIS & $3500$ -- $6800$\,\AA\ & $5$ -- $10$\,\AA\ & {0.007} & 152 & LBG,LAE & (5) \\
Erb+14  & Keck/MOSFIRE & $H$,$K$ & $4$ -- $6$\,\AA\ & {$0.0001$} & 19 & LAE & (6) \\
Matsuda+05 & Subaru/FOCAS & $4700$ -- $9400$\,\AA\ & $10$\,\AA\ & {0.008} & 64 & LAE,LAB & (7) \\
Matsuda+06 & Keck/DEIMOS & $4500$ -- $7500$\,\AA\ & $2.2$\,\AA\ & {0.003} & 39 & LAE,LAB & (8) \\
Yamada+12 & Subaru/FOCAS & $4850$ -- $5100$\,\AA\ & $3$\,\AA\ & {0.004} & 91 & LAE,LAB & (9) \\
{Yamanaka+20} & {Keck/MOSFIRE} & {$K$} & {6\,\AA} & $< 0.001$ & {14} & {LyC,LBG,LAE} & {(10)} \\
Kubo+15 & Subaru/MOIRCS & $H$,$K$ & $10,40$\,\AA\ & {$< 0.001$}  & 38 & DRG,HERO,SMG,XR, & (11) \\
 &  &  &  & & & M24,LAB,LAE,Kdet &  \\
Kubo+16 & Subaru/MOIRCS & $H$,$K$ & $10,40$\,\AA\ & {0.0002} & 3 & SMG & (12) \\
 & Keck/NIRSPEC & $H$,$K$ & $19$\,\AA\ & {0.0003} & 1 & LAB & (12),(13) \\
 & LBT/LUCIFER & $H$,$K$ & $17$\,\AA\ & {0.0003} & 1 & LAB & (12),(13) \\
  & WHT/SAURON & $4810$ -- $5350$\,\AA\ & $4.2$\,\AA\ & {0.001} & 1 & LAB & (12),(14) \\
ADF22 & ALMA & Band\,3 ($3.6$\,mm) & --- & {$< 0.001$} & {12} & SMG & (15) \\
 & Subaru/MOIRCS & $H, K$ & $10, 40$\,\AA\ & {0.0009} & {2} & SMG & (11),(15) \\
 & IRAM/PdBI & Band\,1 ($3.6$\,mm) & --- & {0.0007} & {1} & SMG & (16), (17) \\
 & Keck/MOSFIRE & $K$ & 4 -- 6\,\AA\ & $< 0.001$ & {4} & SMG & (18) \\
Chapman+05 & Keck/ESI & $3200$ -- $10,000$\,\AA\ & $1$\,\AA\ & {0.0008} & 1 & SMG & (19),(20) \\
 & Keck/LRIS & $3100$ -- $8000$\,\AA\ & $5$ -- $8$\,\AA\ & {0.005} & 9 & SMG & (19) \\
Chapman+04 & Keck/LRIS & $3000$ -- $8000$\,\AA\ & $5$ -- $8$\,\AA\ & {0.005} & 2 & OFRG & (21) \\
Saez+15 & Keck/DEIMOS & $4600$ -- $9700$\,\AA\ & $4$\,\AA\ & {0.001} & {89} & XR,bright,LBG,{AGN,Star} & (22) \\
 & Keck/LRIS & $3500$ -- $10,000$\,\AA\ & $10$\,\AA\ & {0.001} & 16 & XR,bright,LBG,{AGN} & (22) \\
 & VLT/VIMOS & $3700$ -- $6700$\,\AA\ & $2.3$\,\AA\ & {0.001} & {277} & XR,bright,LBG,{AGN, Star} & (22) \\
VVDS-Wide & VLT/VIMOS & $5500$ -- $9350$\,\AA\ & $30$\,\AA\ & {0.02} & 912 & bright,AGN & (23),(24) \\
\hline
\end{tabular}
\end{center}
\tablenotetext{a}{Typical FWHM values of spectral resolution}
\tablenotetext{b}{{Uncertainties associated with the systemic redshifts ($z_{\rm sys}$). Typical values at $z \sim 3$ described in the original literature or references therein are listed for VIMOS08, Erb+14, Yamanaka+20, Kubo+15, Kubo+16, Saez+15, and ADF22. For SSA22-HIT, VIMOS06, DEIMOS08, Chapman+05, Chapman+04, and VVDS-Wide where the systemic redshifts are determined based on the cross-correlation technique or based on comparison with spectral templates, we take the spectral resolution as conservative estimates of the redshift uncertainties assuming $z = 3$. For VIMOS12, oIMACS, Steidel+03, Nestor+13, Matsuda+05, Matsuda+06, and Yamada+12, where we converted their $z_{\rm em}$ and/or $z_{\rm abs}$ to $z_{\rm sys}$, we calculated the root-sum-squares of the $z_{\rm em}$/$z_{\rm abs}$ uncertainties and the RMS scatter of the \citet{Adelberger+05} formulae. The $z_{\rm em}$/$z_{\rm abs}$ uncertainties are given by the original literature for Steidel+03 and the spectral resolutions assuming $z = 3$ for the others. } }
\tablenotetext{c}{Number of objects whose redshifts are confirmed in the SSA22-Sb1 field ($334.1014 < {\rm R.A. [deg]} < 334.6532$, $0.0402 < {\rm Decl. [deg]} < 0.4942$). Multiple counts of objects by different catalogs or different instruments are allowed. }
\tablenotetext{d}{(1) \citet{Kousai11} (2) \citet{Hayashino+19} (3) H. Umehata et al. (in preparation) (4) \citet{Steidel+03} (5) \citet{Nestor+13} (6) \citet{Erb+14} (7) \citet{Matsuda+05} (8) \citet{Matsuda+06} (9) \citet{Yamada+12b} (10) \citet{Yamanaka+20} (11) \citet{Kubo+15} (12) \citet{Kubo+16} (13) \citet{McLinden+13} (14) \citet{Weijmans+10} (15) \citet{Umehata+19} (16) \citet{Umehata+17b} (17) \citet{Bothwell+13} (18) \citet{Umehata+18} (19) \citet{Chapman+05} (20) \citet{Chapman+02c} (21) \citet{Chapman+04} (22) \citet{Saez+15} (23) \citet{LeFevre+13} (24) \citet{Garilli+08}}
\tablenotetext{e}{In the catalog compilation, the less reliable redshift measurements with the redshift quality flag $=$ C in SSA22-HIT (Section~\ref{sec:zid_HIT}), VIMOS06, and DEIMOS08 (Section~\ref{sec:otherdata}) are not used. }
\end{table*}

We collected the redshift catalogs available in the SSA22-Sb1 field, which are summarized in Table~\ref{tb:speczcomp}. 
Numerous published works have conducted follow-up spectroscopy of rest-frame UV-selected star-forming galaxies. 
\citet{Hayashino+19} reported 82 LBGs and AGNs at $2 < z < 4$ in the SSA22-Sb1 field, which was based on a deep spectroscopic survey performed with VLT/VIMOS in 2008 (VIMOS08). They measured redshifts in two ways using the Ly$\alpha$ emission line ($z_{\rm em}$) and multiple interstellar absorption lines ($z_{\rm abs}$). Because both often offset from the systemic redshift ($z_{\rm sys}$) owing to the outflowing interstellar gas (e.g., \citealt{Pettini+01,Shapley+03}), they estimated the systemic redshifts by adopting the calibration formulae proposed by \citet{Adelberger+05}. The calibration formulae from $z_{\rm em}$ and/or $z_{\rm abs}$ to $z_{\rm sys}$ were obtained by linear fitting to the measurements of the three types of redshifts for 138 galaxies at $2 < z < 3.5$, where the systemic redshifts were precisely determined through near-infrared (NIR) observations of rest-frame optical nebular emission lines \citep{Adelberger+05}. 
\citet{Steidel+03} conducted a large spectroscopic survey of LBGs using the Keck Low Resolution Imaging Spectrometer (LRIS; \citealt{Oke+95}). They released 940 spectroscopic redshifts at $2 < z < 4$, of which 133 lie in the SSA22-Sb1 field. { Two QSOs were found in the SSA22-Sb1 field.} Their catalog contains two types of redshifts, $z_{\rm em}$ and $z_{\rm abs}$, from which we estimated the systemic redshifts by adopting the calibration formulae of \citet{Adelberger+05}. 
\citet{Nestor+13} confirmed redshifts for 51 LBGs at $2.4 < z < 3.4$ and 101 LAEs at $z \approx 3.1$ using Keck LRIS. Both or either of $z_{\rm em}$ and $z_{\rm abs}$ are contained in their catalog, from which we estimated $z_{\rm sys}$ by adopting the calibration formulae of \citet{Adelberger+05}. 
\citet{Erb+14} used the Multi-Object Spectrometer For Infra-Red Exploration on the Keck telescope (MOSFIRE; \citealt{McLean+10,McLean+12}) to confirm redshifts of 19 LAEs in the $z \approx 3.1$ SSA22 protocluster. While their targets overlap with those of \citet{Steidel+03} and \citet{Nestor+13}, their systemic redshifts determined from the rest-frame optical nebular emission lines ([O\,{\sc iii}]~5007 and H$\alpha$) are more reliable, and no extra calibration is applied. 
\citet{Matsuda+05}, \citet{Matsuda+06}, and \citet{Yamada+12b} confirmed redshifts for 64, 39, and 91 LAEs at $z \approx 3.1$ using the Subaru Faint Object Camera and Spectrograph (FOCAS; \citealt{Kashikawa+02}) and Keck DEIMOS, some of which are LABs. Their redshifts are all determined from the peaks of the Ly$\alpha$ emission lines, from which we estimated $z_{\rm sys}$ by adopting the calibration formulae of \citet{Adelberger+05}. 
\citet{Yamanaka+20} conducted spectroscopic observations with the Keck MOSFIRE-targeting LyCs \citep{Iwata+09,Micheva+17b}, LBGs (this work), and LAEs \citep{Yamada+12a,Yamada+12b}. They detected multiple lines among [O\,{\sc iii}]\,5007, [O\,{\sc iii}]\,4959, and H~{$\beta$} for the two LyCs, 10 LBGs, and one LAE, which provide accurate systemic redshifts. For our catalog compilation, we also took an LBG, SSA22-LBG-05 in \citet{Yamanaka+20}, for which only [O\,{\sc iii}]\,5007 is detected in the MOSFIRE observations, but Ly$\alpha$ is identified in our SSA22-HIT survey. 

In the SSA22 field, galaxies selected at NIR or longer wavelengths have also been actively investigated. They are massive quenched or dust-obscured galaxies at $z \gtrsim 2$. 
\citet{Kubo+15} used Subaru MOIRCS to observe candidate PCs with $K_s < 24$\,mag and $2.6 <$ photo-$z$ $< 3.6$ (Kdet), distant red galaxies (DRGs; $J - K >1.4$; \citealt{vanDokkum+03}), hyper extremely red objects (HEROs; $J - K > 2.1$; \citealt{Totani+01}), Spitzer MIPS $24\,\mu$m sources (M24), AzTEC/ASTE $1.1$\,mm sources (SMGs), Chandra X-ray sources (XRs), LAEs, and LABs. They confirmed systemic redshifts of 39 galaxies at $2.0 < z < 3.4$ from their rest-frame optical nebular emission lines. We ignore J221737.3+001816.0, whose redshift was determined from the Balmer break with large uncertainty. 
In \citet{Kubo+16}, the authors intensively investigated groups of massive galaxies associated with an SMG and LABs at $z = 3.1$ through their NIR spectroscopic observations and archival redshift catalogs. We catalog the six galaxies from \citet{Kubo+16}, avoiding duplication of the source catalogs with the other archival data used in this study. 
The ALMA deep field in the SSA22 (ADF22) survey is a deep $20$\,arcmin$^2$ survey at $1.1$\,mm using ALMA Band 6 \citep{Umehata+17b,Umehata+18}. Among the 35 ALMA-detected SMGs, {21 were spectroscopically confirmed at $2 \lesssim z \lesssim 3$ by the ALMA observations and archival redshift catalogs. We cataloged the 19 galaxies whose redshifts originally come from \citet{Bothwell+13} and \citet{Umehata+15,Umehata+18,Umehata+19}, avoiding duplication of the source catalogs with the other archival data used in this study. Their redshift identifications were based on the NIR or FIR observations of the rest-frame optical or longer-wavelength emission lines, which ensures systemic redshifts.} 
\citet{Chapman+05} performed a redshift survey of SMGs using the Keck/LRIS and Echellette Spectrograph and Imager (ESI; \citealt{Sheinis+02}). Spectroscopic redshifts were confirmed for 73 SMGs, of which 10 were in the SSA22-Sb1 field. The SMGs of \citet{Chapman+05} were originally selected using the Submillimeter Common-User Bolometer Array-2 \citep{Holland+13} equipped with the James Clerk Maxwell Telescope, which has a large beam size of $\gtrsim 10$\arcsec. They attempted to accurately identify the optical counterparts for spectroscopic observations by cross-matching with radio sources in the finer-resolution Karl G. Jansky Very Large Array (VLA) 20\,cm map, but this identification method is not complete \citep{Hodge+13}.  
\citet{Chapman+04} measured spectroscopic redshifts for 18 optical-faint, submillimeter-faint, and radio-bright galaxies (OFRGs) using Keck/LRIS, among which two are in the SSA22-Sb1 field. 
The redshifts in the catalogs of \citet{Chapman+04,Chapman+05} were determined based on a comparison with the template spectra in multiple features such as the Ly$\alpha$ emission line, interstellar absorption lines, and continuum breaks, to which extra calibration was not applied. 

\citet{Saez+15} conducted a large spectroscopic survey of X-ray sources (XRs), bright objects with $R < 22.5$ (bright), and LBGs in the SSA22-Sb1 field using the VLT/VIMOS, Keck/DEIMOS, and LRIS. They confirmed redshifts of the {247 extragalactic sources and 120 Galactic stars} using the same cross-correlation technique as we adopted for the SSA22-HIT data. We did not apply extra calibration to redshift measurements. They adopted their AGN criteria for XR sources spectroscopically confirmed by their observations as well as by \citet{Lehmer+09b}, identifying a total of 84 AGNs. 
The VIMOS VLT Deep Survey in its Wide layer (VVDS-Wide; \citealt{Garilli+08,LeFevre+13}) confirmed $\sim 26,000$ bright galaxies and AGNs with $17.5 < i < 22.5$. We extracted 912 objects in the SSA22-Sb1 field from the public catalog\footnote{\url{https://cesam.lam.fr/vvds/index.php}}{, where we removed objects whose redshifts were tentatively determined with their flags of 1, 11, 21, or 211}. {The majority of the 912 objects are at $z < 2$, where only 10 lie beyond $z > 2$ and the median redshift is $z \approx 0.6$. Their redshifts were measured based on the cross-correlation technique, followed by careful inspection by multiple people, to which extra calibration was not applied.} We consider the cataloged objects with any broad ($\gtrsim 1000$\,km\,s$^{-1}$) emission lines resolved at the VIMOS resolution as AGNs{, which were selected by the cataloged flags of $10 < {\rm flag} < 20$ or ${\rm flag} > 200$}. 

We also utilized the following archival redshift catalogs that are not publicly available. 
The VIMOS06 survey \citep{Kousai11,Hayashino+19} was a pilot observation for the VIMOS08 survey using VLT/VIMOS with LR-blue grism under the program ID 077.A-0787 (PI: R. Yamauchi), where a wider area was surveyed with a shorter integration time than VIMOS08. We performed redshift determination for 98 spectra that were reduced by \citet{Kousai11} and \citet{Hayashino+19} in the same manner as that adopted for the SSA22-HIT data. Eventually, we confirm redshifts for 70 LBGs and six AGNs, where the AGN identification is based on \citet{Steidel+03} and \citet{Hayashino+19}. 
H. Umehata et al. (2023, in preparation) conducted a spectroscopic survey of SMGs, LBGs, and LyCs using VLT/VIMOS with LR-blue grism (VIMOS12) under the program ID 089.A-0740 (PI: H. Umehata). They measured both $z_{\rm em}$ and $z_{\rm abs}$, from which we estimated the systemic redshifts by adopting the calibration formulae of \citet{Adelberger+05}. 
Using Keck/DEIMOS with the 900ZD grating under the program ID of S275D, T. Yamada et al. performed spectroscopic observations (DEIMOS08) of LAEs, LBGs, Ly$\alpha$ absorbers (LAAs; Hayashino et al. 2004), and $K_s$-detected objects (Kdet) not only in the SSA22-Sb1 field but also in a wider area. We conducted reduction and redshift identification in the same manner as that adopted for the SSA22-HIT data. Out of 94 galaxies whose redshifts were confirmed, 42 lie in the SSA22-Sb1 field. 
M. Ouchi et al. conducted a large spectroscopic survey of high-$z$ galaxies using the Inamori Magellan Areal Camera and Spectrograph (IMACS; \citealt{Dressler+11}) on the Magellan I Baade Telescope between 2007 and 2010 (oIMACS; \citealt{Momcheva+13,Higuchi+19}). We analyzed the spectra of the objects in the SSA22 field that were obtained using Gri-300-17.5 and Gri-300-4.3 grisms. Redshifts were determined from their Ly$\alpha$ emission lines, from which we estimated the systemic redshifts by adopting the calibration formulae of \citet{Adelberger+05}. In this work, we used 29 LAEs, LABs, LBGs, and LyCs in the SSA22-Sb1 field. 


The determination schemes for the above redshifts were divided into three types. The redshift determination based on the NIR or FIR observations of the rest-frame optical or longer-wavelength emission lines yielded the most reliable systemic redshifts (\citealt{Erb+14,Kubo+15,Kubo+16,Yamanaka+20}; ADF22). The cross-correlation or comparison with spectral templates, which was adopted in SSA22-HIT, \citet{Chapman+04}, \citet{Chapman+05}, \citet{Saez+15}, VVDS-Wide, VIMOS06, and DEIMOS08, is expected to provide systemic redshifts when the spectral properties of the used templates are similar to those of the observed objects. The empirical calibration formulae from $z_{\rm em}$ and/or $z_{\rm abs}$ to $z_{\rm sys}$, which were applied to the data of VIMOS08, \citet{Steidel+03}, \citet{Nestor+13}, \citet{Matsuda+05}, \citet{Matsuda+06}, \citet{Yamada+12b}, VIMOS12, and oIMACS, are reliable within $\sigma_{z} \approx 0.003$ for star-forming galaxies at $2 < z < 3.5$ \citep{Adelberger+05}. In Table~\ref{tb:speczcomp}, we list the uncertainties of the systemic redshifts in the individual works. If uncertainties are described in the original literature or references therein, we use them only. Otherwise, we conservatively expect the redshift errors to be $(1 + z) \Delta \lambda / \lambda = (1 + z) / R$, where $\Delta \lambda$ and $R$ correspond to the spectral resolutions in the individual works. If the calibration formulae of \citet{Adelberger+05} are adopted, we include the associated uncertainty ($\sigma_{z} \approx 0.003$) by taking the root sum squares.

\subsection{Catalog Compilation} \label{sec:compcat}

\begin{figure*}
\begin{center}
\includegraphics[width=1.0\linewidth, angle=0]{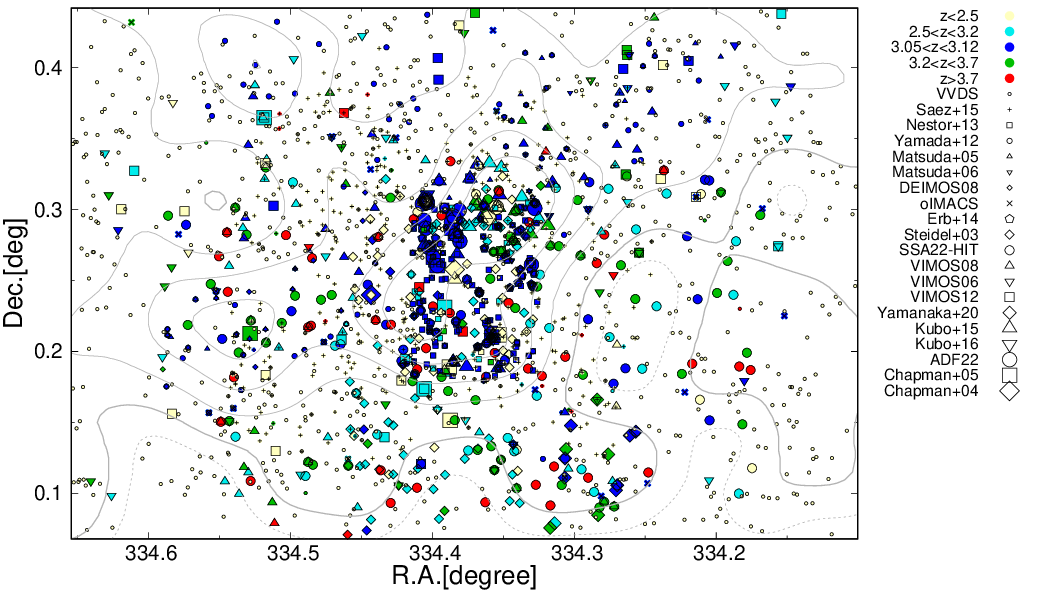}
\caption{Sky distribution of objects with spectroscopically confirmed redshifts in the SSA22-Sb1 field. The symbol difference corresponds to the references (see the text and Table~\ref{tb:speczcomp}). Redshifts are expressed by the following symbols: $z < 2.5$ in light yellow, $2.5 \leq z < 3.2$ in cyan, $3.05 \leq z \leq 3.12$ in blue, $3.2 \leq z < 3.7$ in green, and $z \geq 3.7$ in red. The background contours show the $z = 3.1$ LAE number density, which are the same as those in Figure~\ref{fig:skydist_targets}.  \label{fig:skydist_comp}}
\end{center}
\end{figure*}

\begin{figure*}
\begin{center}
\includegraphics[width=1.0\linewidth, angle=0]{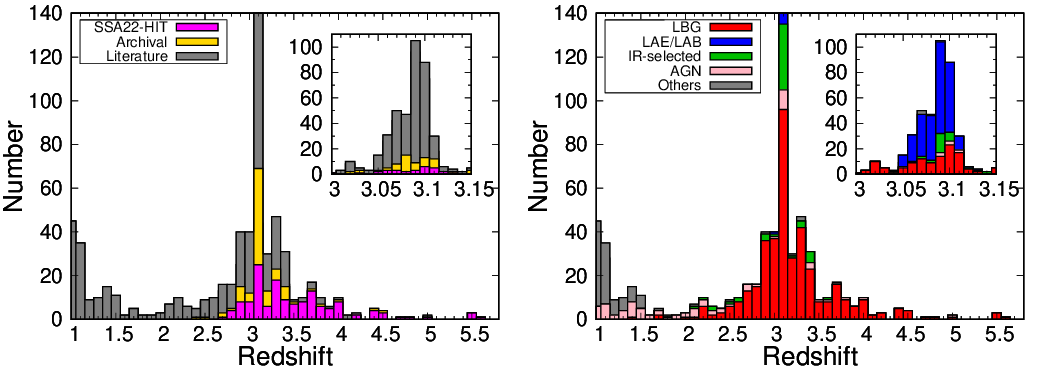}
\caption{Redshift distribution of the objects spectroscopically identified in the SSA22-Sb1 field. In the left panel, the objects are separated by their references (see the text and Table~\ref{tb:speczcomp}): the SSA22-HIT sample in magenta, {archival} sample (VIMOS06, VIMOS12, DEIMOS08, and oIMACS) in yellow, and other published samples in gray. In the right panel, the objects are separated by their object type: LBGs in red, LAEs/LABs in blue, IR-selected galaxies (Kdet, DRG, HERO, M24, SMG, and OFRG) in green, {AGNs in light-pink}, and the others in gray. For display purposes, the vertical axis value is limited to 140; however, it hides $\sim 250$ objects at $z \approx 3.1$. We then embedded small subpanels showing finely binned histograms around $z = 3.1$. \label{fig:zdist_comp}}
\end{center}
\end{figure*}

We merged our SSA22-HIT redshift catalog (Section~\ref{sec:zid_HIT}) and ancillary catalogs (Section~\ref{sec:otherdata}). We here excluded 23 objects with the redshift quality flag $=$ C (possible redshift cases) in SSA22-HIT to match the redshift quality with the other catalogs. For the same reason, the objects with the flag $=$ C were also excluded from the VIMOS06 and DEIMOS08 samples whose redshifts were determined in the same manner as that adopted for the SSA22-HIT data (Section~\ref{sec:otherdata}). 
The individual catalogs follow their own astrometry, which is slightly offset by $\sim 1$\arcsec at maximum. We first searched for counterparts and extracted their coordinates {in the $V$ image for the LBGs in the \citet{Steidel+03} and \citet{Nestor+13} catalogs; the NB497 image for the LAEs; the $K_s$ image for galaxies selected at NIR or longer wavelengths in \citet{Kubo+15}, \citet{Kubo+16}, ADF22 catalog, and the $i'$ image for the other catalog objects.} 

We found 228 objects duplicated in multiple catalogs with a matching radius of $1.0$\arcsec, which yields 490 pairs of redshifts measured by different works. We estimated the average and standard deviation of the redshift differences to be $\approx -0.001$ and $\approx 0.003$, respectively. The redshifts in our compilation are roughly consistent with each other within the uncertainties (see Section~\ref{sec:otherdata} and Table~\ref{tb:speczcomp}). On the other hand, there are 32 outlier pairs for 21 objects with redshift differences as large as $\Delta z > 0.02$. Approximately one-third of them are due to the misidentification of their emission lines (e.g., Ly$\alpha$ at $z \approx 3.1$ and [O\,{\sc ii}]\,3727 at $z \approx 0.33$). For the duplicated objects, representative redshifts are selected according to the following order of priority: \citet{Erb+14}, \citet{Yamanaka+20}, ADF22, \citet{Kubo+15}, \citet{Kubo+16}, \citet{Chapman+04}, \citet{Chapman+05}, \citet{Steidel+03}, \citet{Nestor+13}, SSA22-HIT, VIMOS08, VIMOS12, VIMOS06, \citet{Matsuda+06}, \citet{Matsuda+05}, \citet{Yamada+12a}, DEIMOS08, VVDS, \citet{Saez+15}, and oIMACS. This priority order is determined by whether the redshift determination is based on the rest-frame optical nebular emission lines, the survey depth, the spectral resolution, and/or the details of the classification of the object types.

We isolated AGNs by matching our compiled catalogs with AGN catalogs. \citet{Saez+15} performed secure AGN selection for Chandra XR sources of \citet{Lehmer+09a,Lehmer+09b} based on the X-ray luminosity, spectral shape, X-to-optical, and X-to-radio flux ratios. \citet{Hayashino+19} selected AGNs based on visual inspection of the optical spectra. \citet{Micheva+17a} searched for LyC leakers among 14 AGNs at $z \geq 3.06$ selected by detection in X-ray \citep{Lehmer+09a,Saez+15} or by emission lines in the optical spectra. We also visually inspected the SSA22-HIT spectra to identify four AGNs with strong rest-UV emission lines, namely, C\,{\sc iv} ($1548.20$\,\AA\ and $1550.78$\,\AA), He\,{\sc ii} ($1640.4$\,\AA), and C\,{\sc iii}] ($1906.68$\,\AA\ and $1908.73$\,\AA). We eventually found 105 AGNs, including 34 at $z \geq 2$. 

We compiled a final catalog of $1941$ unique objects with spectroscopic redshifts in the SSA22-Sb1 field, among which $730$ are at $z \geq 2$. The sky distribution is shown in Figure~\ref{fig:skydist_comp}. The redshift histograms are shown in Figure~\ref{fig:zdist_comp}, where representative redshifts were used. The number of galaxies at $z > 3.2$, which can be used as background light sources for the H\,{\sc i} tomography at $z = 3.1$ or more, is significantly increased by a factor of $1.7$ owing to the SSA22-HIT survey. We have made the compiled redshift catalog public (see Appendix~\ref{sec:ap_pubcat}). 
The SSA22 field is one of the most extensively investigated regions with various types of galaxies. Our compiled redshift catalog is useful for a variety of applications. For example, one of our future works will be to investigate the dependency of the different types of galaxies on the IGM H\,{\sc i} environment, which will be complementary to the study in a general field by \citet{Momose+21a}. 

\section{Comparison with Other High-$z$ Fields}\label{sec:comp_survey}

\begin{figure}
\begin{center}
\includegraphics[width=1.0\linewidth, angle=0]{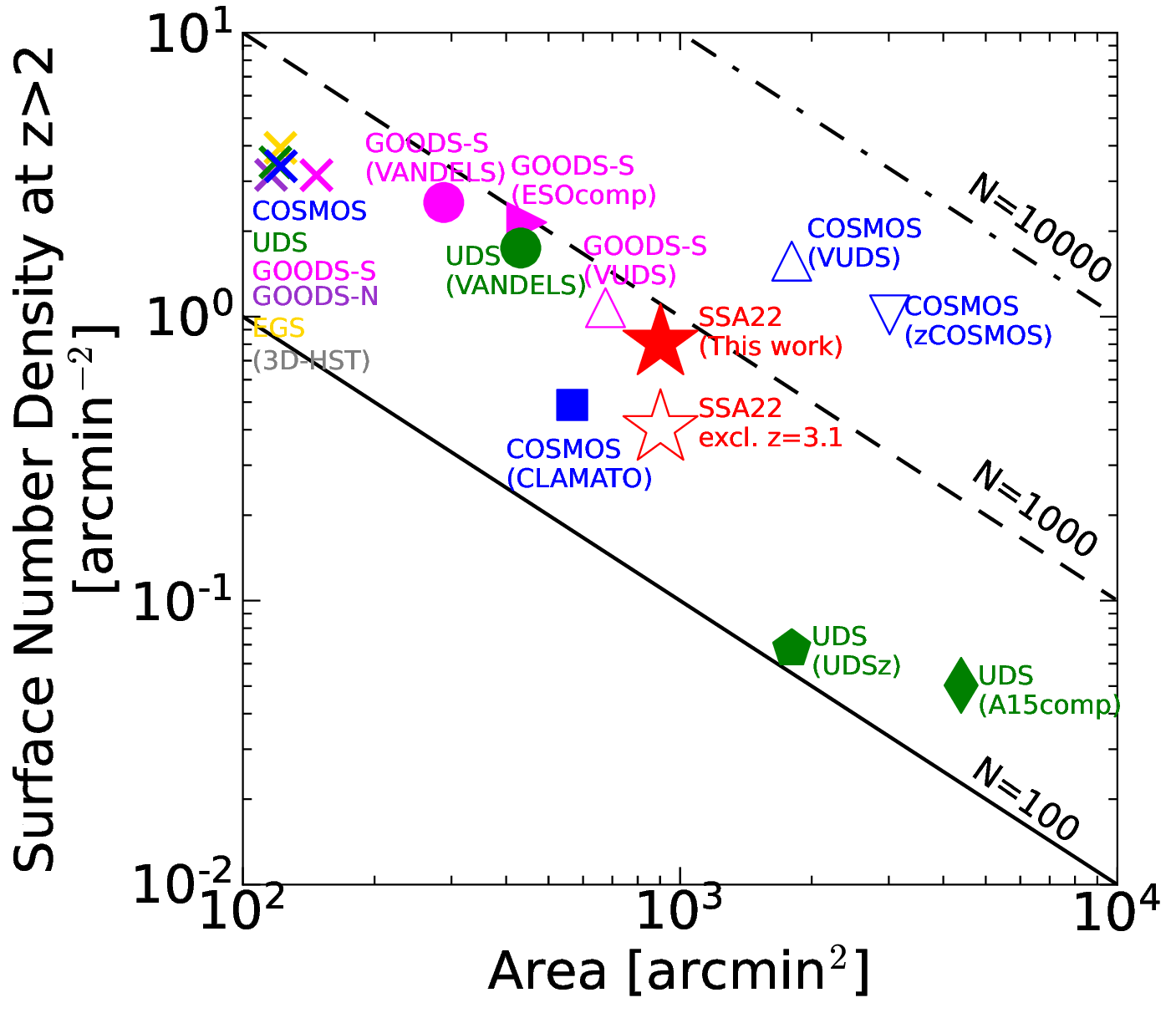}
\caption{Comparison of the covered area and the surface number density of objects at $z > 2$ in the SSA22-Sb1 compiled catalog with other published redshift catalogs available in the major high-$z$ survey fields (see the text). Red filled/open stars correspond to the SSA22-Sb1 compiled catalog including/excluding PC galaxies at $3.05 \leq z \leq 3.12$. Differences in the symbol colors represent the differences in the fields: blue, green, magenta, violet, and yellow symbols for COSMOS, UDS, GOODS-S, GOODS-N, and EGS fields, respectively. Different symbol types represent surveys/projects from which the catalogs are constructed: crosses, open inverted triangles, open triangles, pentagons, circles, squares, right-pointing triangles, and diamonds correspond to 3D-HST \citep{Momcheva+16}, zCOSMOS \citep{Lilly+07,Mignoli+19}, VUDS \citep{LeFevre+15}, UDSz \citep{Bradshaw+13,McLure+13b}, VANDELS \citep{Garilli+21}, CLAMATO \citep{KGLee+18}, ESOcomp \citep{Popesso+09,Baldassare+15}, and A15comp \citep{Smail+08,Simpson+12,Akiyama+15}, respectively. The solid, dashed and dotted-dashed lines correspond to the total number of objects at $z > 2$ equal to 100, 1,000, and 10,000, respectively.  \label{fig:CompSurvey_numden}}
\end{center}
\end{figure}

\begin{figure}
\begin{center}
\includegraphics[width=1.0\linewidth, angle=0]{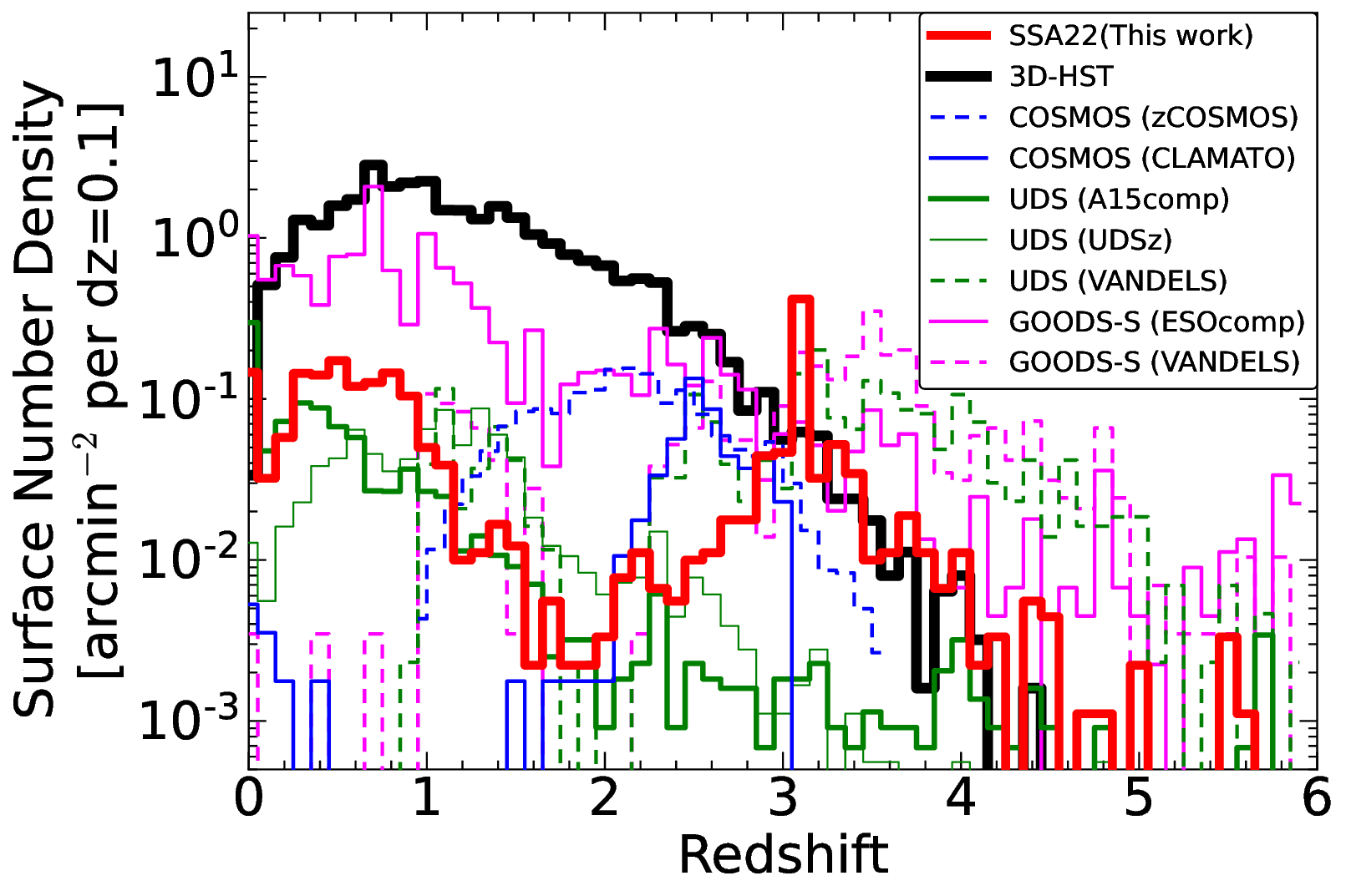}
\caption{Comparison of the surface number density per $\Delta z = 0.1$ of objects in the SSA22-Sb1 compiled catalog with other published redshift catalogs available in the major high-$z$ survey fields (see the text). The red solid, blue dashed, blue solid, green thick, green thin, and magenta solid histograms correspond to the redshift distribution of the SSA22 compilation (this work), zCOSMOS \citep{Lilly+07,Mignoli+19}, CLAMATO \citep{KGLee+18}, A15comp \citep{Smail+08,Simpson+12,Akiyama+15}, UDSz \citep{Bradshaw+13,McLure+13b}, and ESOcomp \citep{Popesso+09,Baldassare+15}, respectively. For the 3D-HST catalog, data from the five fields were plotted together using the black bold histogram. Green and magenta dashed histograms are obtained from VANDELS \citep{Garilli+21} but in UDS and GOODS-S, respectively. \label{fig:CompSurvey_zdist}}
\end{center}
\end{figure}

There are a limited number of fields where rich multiband imaging and spectroscopic data are available and are sufficiently deep to confirm galaxies at high redshifts, namely $z \geq 2$. Here, we selected five major fields for high-$z$ galaxy investigation: the Cosmic Evolution Survey (COSMOS; \citealt{Scoville+07}), the UKIRT Infrared Deep Sky Survey (UKIDSS) Ultra-Deep Survey (UDS; \citealt{Lawrence+07}), the Extended Groth Strip (EGS; \citealt{Davis+07}), and the Great Observatories Origin Deep Survey South and North (GOODS-S and -N; \citealt{Giavalisco+04}). It is worth comparing our compilation catalog of spectroscopic redshifts with published redshift catalogs in the five major high-$z$ fields to characterize the SSA22-Sb1 field. 

We collected the following catalogs available in the five fields. 
The 3D-HST catalog \citep{Momcheva+16} is available for all five fields. Most of their redshifts are determined from a combination of the $HST$ grism spectra and multiband photometry, which have relatively large uncertainties. For a small number of galaxies, redshifts were obtained from ground-based spectroscopic surveys. While the 3D-HST catalog yields redshifts with a high surface number density of galaxies, the covered area is limited to $\sim 120$\,arcmin$^2$ per field. 

In the COSMOS field, wider areas, $\sim 3000$, $\sim 1800$, and $\sim 570$\,arcmin$^2$, were spectroscopically observed in the zCOSMOS-Deep \citep{Lilly+07}, VIMOS Ultra-Deep Survey (VUDS; \citealt{LeFevre+15}), and CLAMATO (DR1; \citealt{KGLee+18}) projects, respectively. While the zCOSMOS-Deep and VUDS catalogs are not or partially available on their websites, we extracted information about the numbers of spectroscopically confirmed objects and their redshift distributions from the literature \citep{LeFevre+15,Mignoli+19}. They are rough estimates but are sufficient for comparison in this study. 

In the UDS field, redshift catalogs are available from UDSz \citep{Bradshaw+13,McLure+13b}, the compilation project led by Akiyama, Simpson, Croom, Geach, Smail, and van Breukelen (hereafter, A15comp; \citealt{Smail+08,Simpson+12,Akiyama+15}), and VANDELS \citep{Garilli+21}. Their covered areas are $\sim 1800$, $\sim 4400$, and $\sim 430$\,arcmin$^2$ for UDSz, A15comp, and VANDELS, respectively. 

In the GOODS-S field, spectroscopic redshift measurements are available from the ESO compilation project \footnote{\url{https://www.eso.org/sci/activities/garching/projects/goods/MasterSpectroscopy.html}} (hereafter, ESOcomp; \citealt{Popesso+09,Balestra+10}), VUDS, and VANDELS. They cover areas of $290$ -- $680$\,arcmin$^2$ that are wider than those of the 3D-HST catalog. 


Figure~\ref{fig:CompSurvey_numden} shows a comparison of the above catalogs in the covered area and the surface number density of spectroscopically confirmed objects, where we limit the objects at $z \geq 2$. 
Because many surveys in the SSA22-Sb1 field focus on the $z \approx 3.1$ PC galaxies, we also plotted the SSA22-Sb1 data excluding galaxies at $3.05 \leq z \leq 3.12$ in Figure~\ref{fig:CompSurvey_numden} by an open star. As seen in this figure, the compiled catalog of SSA22-Sb1 is comparable to the catalogs of the major fields in terms of a well-balanced combination of survey area and surface number density of objects with spectroscopic redshifts. 

We also compared the surface number densities of the cataloged objects as a function of the redshift in Figure~\ref{fig:CompSurvey_zdist}. We note that data from the VUDS catalogs in COSMOS and GOODS-S are not shown because the full-area catalogs are not yet available on their websites. The SSA22-Sb1 compiled catalog maintains a higher surface number density at $z \gtrsim 3$ than the zCOSMOS, CLAMATO, UDSz, and A15comp catalogs. While the catalogs from 3D-HST, VANDELS, and ESOcomp are comparable or superior to the SSA22-Sb1 compiled catalog in the surface number density, their covered areas are more than twice as small as that of SSA22-Sb1. These facts suggest that the SSA22-Sb1 field is suitable for investigating the large-scale structure of galaxies and H\,{\sc i} gas (i.e., H\,{\sc i} tomography) at $z \gtrsim 3$.

\section{SSA22-HIT data quality for H\,{\sc i} tomography}\label{sec:HITquality}

\begin{figure}
\begin{center}
\includegraphics[width=1.0\linewidth, angle=0]{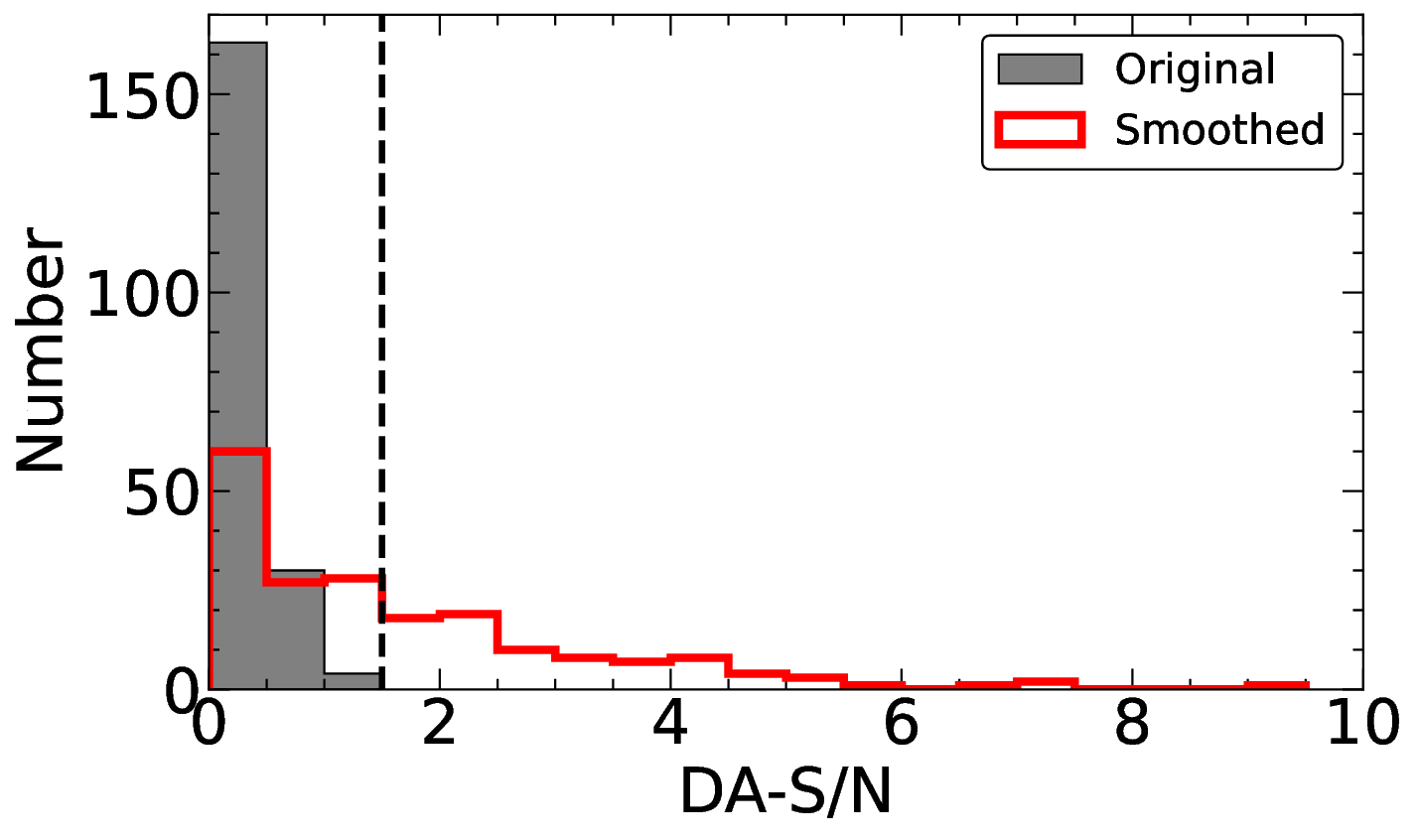}
\caption{DA-S/N distributions for the original (gray filled histogram) and smoothed (red open histogram) SSA22-HIT spectra. In both cases, the QSO (ID 06090) is not shown because its DA-S/N is higher than $10$. Objects with negative DA-S/N are counted in the smallest DA-S/N bin. The vertical dashed line corresponds to DA-S/N $= 1.5$ that is the threshold for the usage in H\,{\sc i} tomography. \label{fig:DASN_all}}
\end{center}
\end{figure}

\begin{figure}
\begin{center}
\includegraphics[width=1.0\linewidth, angle=0]{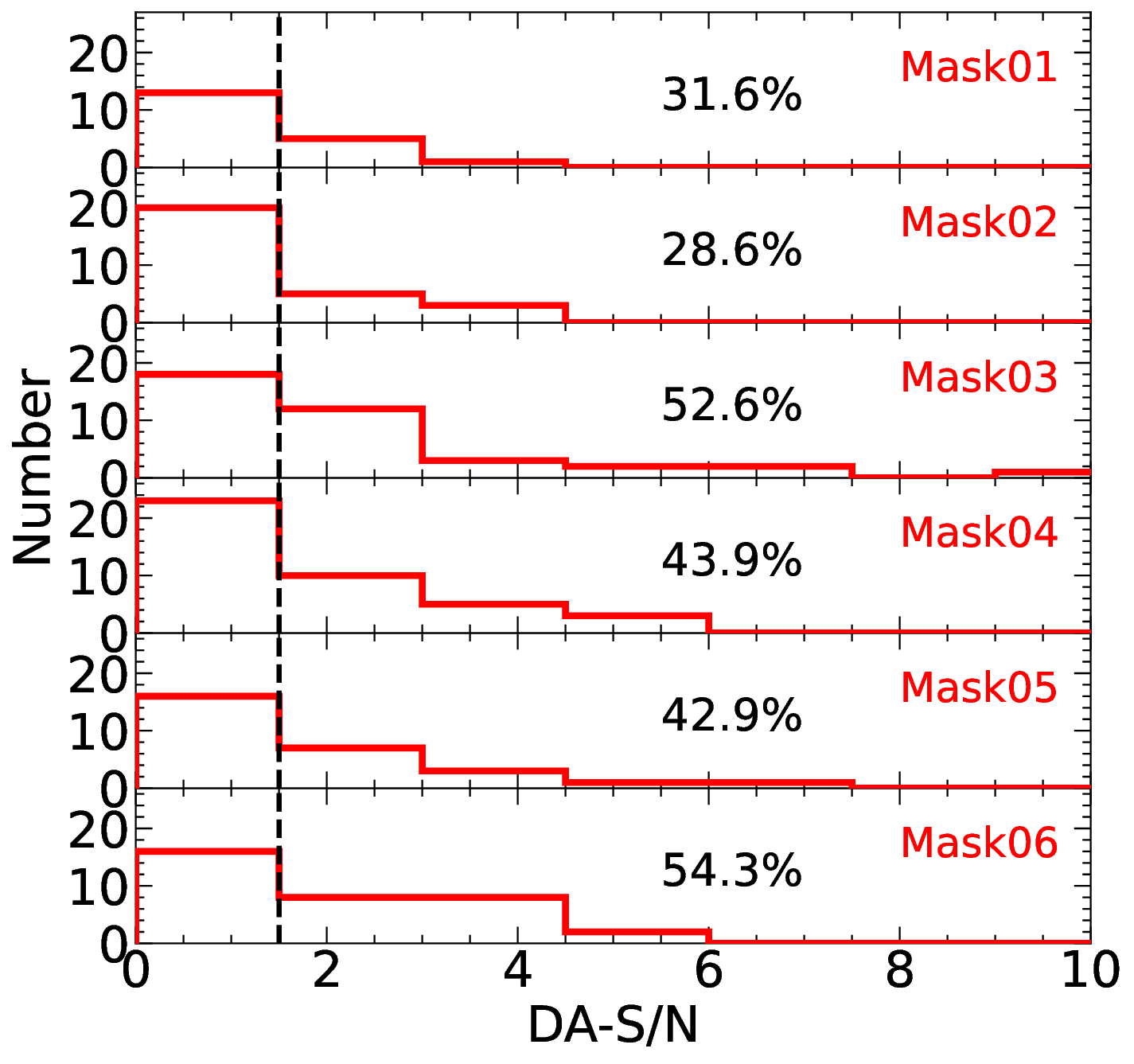}
\caption{DA-S/N distributions for the smoothed SSA22-HIT spectra divided by the DEIMOS mask. The vertical dashed line corresponds to DA-S/N $= 1.5$. The number embedded in each panel indicates the fraction of the objects with DA-S/N $> 1.5$ among ones whose DA ranges are available. \label{fig:DASN_mask}}
\end{center}
\end{figure}

\begin{figure}
\begin{center}
\includegraphics[width=1.0\linewidth, angle=0]{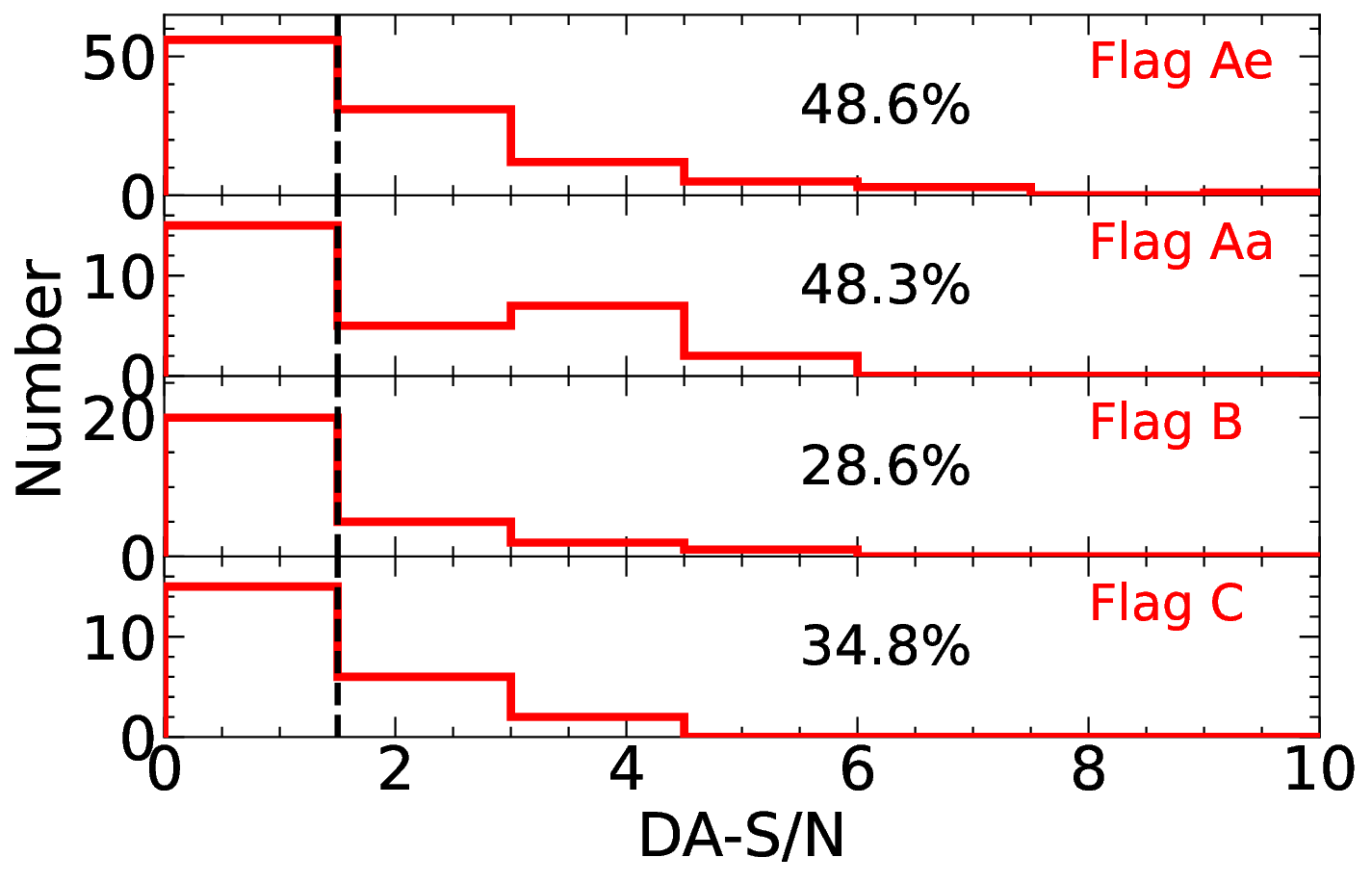}
\caption{Same as Figure~\ref{fig:DASN_mask} but the smoothed SSA22-HIT spectra are divided by the redshift quality flag. For visibility, the number range is set differently for each panel.  \label{fig:DASN_flag}}
\end{center}
\end{figure}

For H\,{\sc i} tomography, not only is the number density of the background light sources important, but so is the quality of their continuum spectra. In this section we evaluate the quality of the SSA22-HIT spectra from the viewpoint of their applicability to the H\,{\sc i} tomographic mapping. 
We here focus on the wavelength range at $1045$\,\AA\ $\leq \lambda \leq$ $1185$\,\AA\ in the rest-frame. This wavelength coverage is almost the same as the so-called ``depression at Ly$\alpha$ (DA)'' range (e.g., \citealt{Madau95, Hayashino+19}) and used for the H\,{\sc i} absorption analysis. We consider that the galaxies whose continuum S/N per pixel averaged in the DA range (DA-S/N) is larger than $1.5$ are useful for the H\,{\sc i} absorption analysis. This criterion is comparable to the previous works \citep{KGLee+18,Mukae+20a}. 

Among the 198 spectroscopically confirmed objects in SSA22-HIT, 189 cover the DA range in their observed spectra. Unfortunately, all of them have DA-S/N lower than 1.5 except for the QSO (ID 06090), as shown in Figure~\ref{fig:DASN_all}. The low DA-S/N is caused by the weather conditions that were not the best, the lower instrumental sensitivity at shorter wavelength (Section~\ref{sec:datareduc}), and the significant Ly$\alpha$ depression for $z > 3$ galaxies (e.g., \citealt{Madau95,Inoue+14}). 
We then smoothed the SSA22-HIT spectra by the Gaussian kernel with a standard deviation of $10.4$\,\AA\ or FWHM (FWHM$_G$) of $24.5$\,\AA\ $\approx 39$ pixels so that their spectral resolutions become similar to those of the VIMOS08 and VIMOS06 observations, namely $\Delta \lambda \sim  25$\,\AA\ in FWHM. The flux density errors are reduced roughly by $1/\sqrt{{\rm FWHM}_G [{\rm pix}]} \sim 1/6$. This smoothing procedure not only improves the DA-S/N but also enables us to combine spectra from SSA22-HIT and VIMOS observations as the background sight-lines in H\,{\sc i} tomography. After the smoothing, 83 out of the 198 spectroscopically confirmed SSA22-HIT galaxies satisfy DA-S/N $> 1.5$. Their DA-S/N distribution is shown in Figure~\ref{fig:DASN_all}. We note that some spectra have negative flux and then negative S/N in the DA range. This is due to over-sky-subtraction caused by contamination of stray light or nearby bright objects in the sky pixels. In Figure~\ref{fig:DASN_all} we put such objects with negative DA-S/N to DA-S/N $= 0$ bin. 

We further investigated the DA-S/N distributions for the smoothed SSA22-HIT spectra of different DEIMOS masks and redshift quality flags, which are shown in Figures~\ref{fig:DASN_mask} and \ref{fig:DASN_flag}, respectively. In the figures, we also show percentages of the objects with DA-S/N $> 1.5$. The percentages in mask01 and mask02 are smaller than those in other masks. This simply reflects their shallower depths (Table~\ref{tb:HITzsum}). There are more objects available as the background light sources in the redshift quality flags A$_{\rm e}$ and A$_{\rm a}$ than in the flag B and C (Figure~\ref{fig:DASN_flag}). We also investigated the DA-S/N distribution for each target category, finding no statistically robust difference. This is due to the small sample size except for the T2 category (see also Table~\ref{tb:HITzsum_T}). Most of the objects with DA-S/N $> 1.5$ come from the T2 category galaxies whose actual redshifts broadly span $3 \lesssim z \lesssim 4$ (Figure~\ref{fig:zdist_HIT}). 

\begin{figure}
\begin{center}
\includegraphics[width=1.0\linewidth, angle=0]{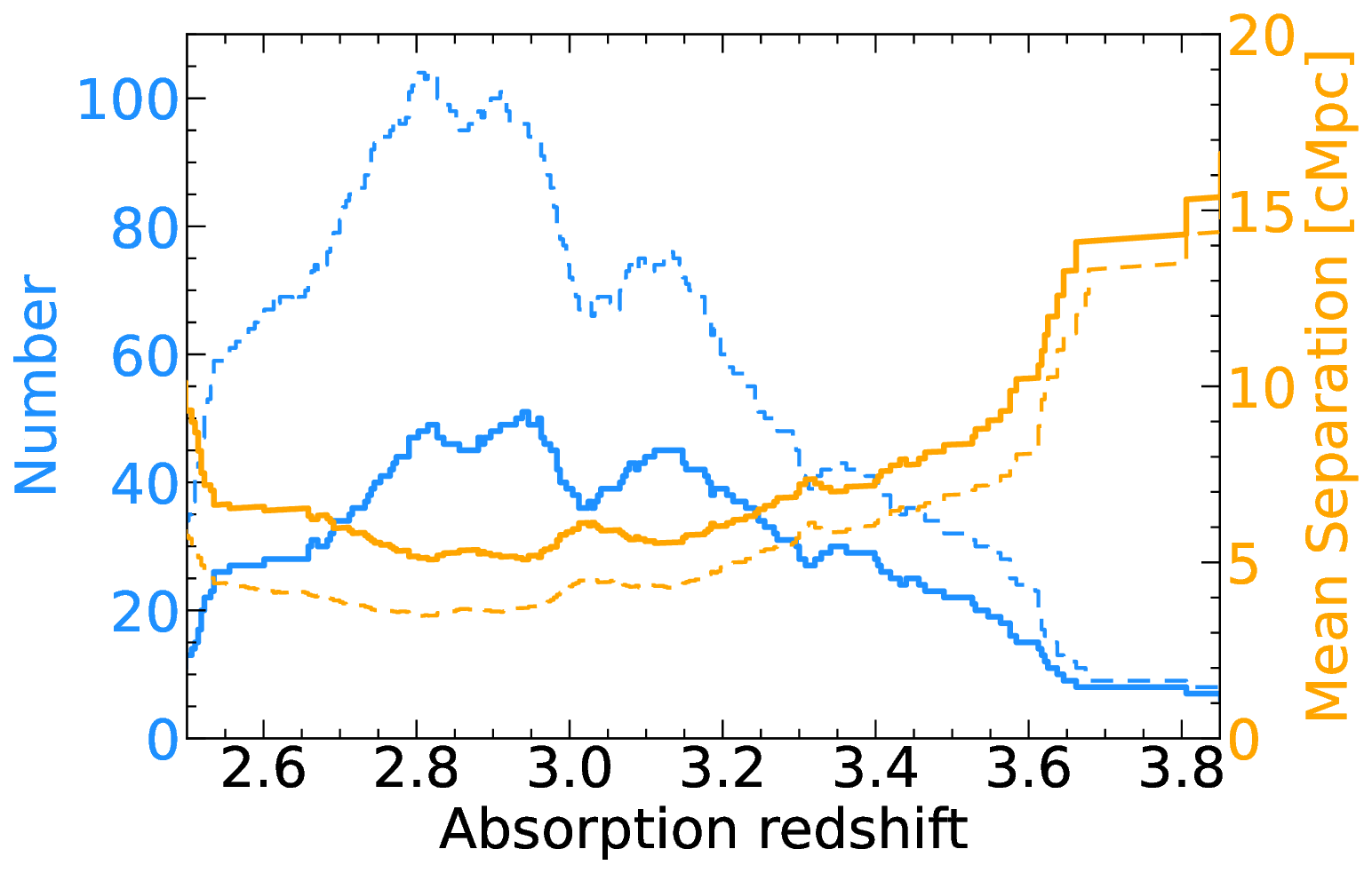}
\caption{Variation of the available background objects (light blue histograms) and the mean separation in cMpc units (orange histograms) as a function of Ly$\alpha$ absorption redshift. The solid histograms are for the smoothed SSA22-HIT sample alone while the dashed histograms are for the combined sample of SSA22-HIT, VIMOS06, and VIMOS08. \label{fig:zdist_LOS}}
\end{center}
\end{figure}

For the individual background objects satisfying DA-S/N $> 1.5$, the H\,{\sc i} Ly$\alpha$ absorption can be measured only in the redshift range corresponding to the DA range. The available redshift range (the Ly$\alpha$ absorption redshift $z_{\rm abs}$) is defined by $z_{\rm abs} = \lambda_{\rm DA} / 1215.67 -1$, where $\lambda_{\rm DA}$ is the DA wavelength range in the observed frame. Figure~\ref{fig:zdist_LOS} shows the number of the available background objects as a function of the Ly$\alpha$ absorption redshift (light blue solid histogram), where we used the smoothed spectra. Using the survey area of $26 \times 15$\,arcmin$^2$, we calculated the surface number density and then mean separation of the sight-line objects. The latter yields a rough estimate of the resolving power for the H\,{\sc i} gas spatial distribution, which is also shown in Figure~\ref{fig:zdist_LOS} (orange solid histogram). At $z \sim 3$, the number histogram has a peak, and the mean separation of the sight-line objects reaches $\sim 6$\,cMpc. This suggests that we can make an H\,{\sc i} tomography map at $z \sim 3$ with the spatial resolution of $\sim 6$\,cMpc using the SSA22-HIT data.  

We here consider adding VIMOS06 and VIMOS08 samples to increase the number density of the background sight-lines for H\,{\sc i} tomography. We found 20 and 57 objects satisfying DA-S/N $> 1.5$ and lying inside the SSA22-HIT survey area from VIMOS06 and VIMOS08, respectively. The number and mean separation of the available sight-lines from the combined sample of SSA22-HIT, VIMOS06, and VIMOS08 are shown in Figure~\ref{fig:zdist_LOS} by the light blue and orange dashed histograms, respectively. By combining the data taken in the different observations, we can improve the spatial resolution in the H~{\sc i} tomography down to $\sim 4$\,cMpc. 

More practically, we need to carefully consider masking the intrinsic metal absorption lines in the DA range, visual inspection of the spectra, which redshift quality flags to use, area used for the mapping, and so on. The detailed procedures in the H\,{\sc i} map making is discussed in Paper~II.

\section{Summary}\label{sec:summary}

In this study, we describe the survey design, observations, and data reduction of the SSA22 H\,{\sc i} Tomography (SSA22-HIT) survey, as well as the creation of a large compilation catalog of spectroscopic redshifts available in the SSA22-Sb1 field. 

In the SSA22-HIT survey, we spent an effective observation time of 41\,h with the Keck telescope DEIMOS to carry out a large spectroscopic campaign of galaxies at $z > 2.5$ in the $26 \times 15$\,arcmin$^2$ area of the SSA22-Sb1 field. We confirmed the spectroscopic redshifts of 198 galaxies, of which 148 were newly determined in this work. The SSA22-HIT survey significantly increases the number of galaxies at $z > 3.2$ by a factor of $1.7$ even after excluding the objects with the lowest redshift quality flag (C). This greatly benefits the investigation of H\,{\sc i} Ly$\alpha$ absorption at the protocluster redshift $z = 3.1$ or higher. 

We merged our redshift measurements with those of previous studies, where the possible measurements in SSA22-HIT (the redshift quality flag $=$ C) are not used. The compiled catalog, which lists not only the spectroscopic redshifts but also the object types, includes $\sim 1940$ and $\sim 730$ objects in total and at $z \geq 2$, respectively. Because such a catalog is useful in a variety of applications, we have made it public (Appendix~\ref{sec:ap_pubcat}). We compared our compiled redshift catalog in the SSA22-Sb1 field with publicly available catalogs in the five major extragalactic fields. In the parameter space of the covered area and the surface number density of objects at $z \geq 2$, the SSA22-Sb1 catalog is comparable to the other field catalogs. Especially at $z \gtrsim 3$, the SSA22-Sb1 catalog contains a higher surface number density of objects than the other field catalogs with similar or larger covered areas (zCOSMOS, CLAMATO, UDSz, and A15comp). 

We evaluated S/N in the DA range of the SSA22-HIT spectra for H\,{\sc i} tomography. While a sufficient amount of smoothing is needed to make the spectral resolution worse from $\Delta \lambda \approx 4.7$\,\AA\ to $25$\,\AA, about 40\,\% of the spectroscopically confirmed objects in SSA22-HIT have DA-S/N high enough for analyzing the H\,{\sc i} Ly$\alpha$ absorption. These available sight-line data enable us to construct an H\,{\sc i} tomography map with a spatial resolution $\sim 6$\,cMpc at $z \sim 3$. The mapping resolution can be improved down to $\sim 4$\,cMpc if combining the data taken in SSA22-HIT and the two precedent VIMOS observations. 

From the above, we conclude that our data set in the SSA22-Sb1 field is suitable for investigating large-scale spatial distributions of galaxies and H\,{\sc i} gas, especially at $z \gtrsim 3.1$. In Paper II (K. Mawatari et al. 2023, in preparation), we demonstrate H\,{\sc i} tomography at $z \sim 3$ in the SSA22-Sb1 field containing the prominent protocluster and discuss the spatial alignment between H\,{\sc i} and galaxies. It is also interesting to investigate the dependency of the different types of galaxies on the IGM H\,{\sc i} environment as a future work. 


\acknowledgments
K.M. acknowledges financial support from the Japan Society for the Promotion of Science (JSPS) through KAKENHI grant No. 20K14516. K.M. and A.K.I. are supported by JSPS KAKENHI Grant Nos. 26287034 and 17H01114. H.U. is supported by JSPS Grant-in-Aid for Research Activity Start-up (16H06713) and KAKENHI grant No. 20H01953. This work was partially supported by the joint research program of the Institute for Cosmic Ray Research (ICRR), University of Tokyo. This work is supported by the World Premier International Research Center Initiative (WPI Initiative), MEXT, Japan, as well as KAKENHI Grant-in-Aid for Scientific Research (A) (15H02064 and 19H00697) through the JSPS. 
We thank Kohei Shinoka, Marc Kassis, Joel Aycock, Heather Hershley, and Jim Lyke for their help during the observations with the Keck/DEIMOS. We also thank Hassen M. Yesuf and Mariko Kubo for their advice in the data reduction. 
We thank the anonymous referee for providing constructive comments that significantly improved the clarity of the paper. 
Data analysis was in part carried out on the Multi-wavelength Data Analysis System operated by the Astronomy Data Center (ADC), National Astronomical Observatory of Japan. 
We would like to thank Editage (\url{www.editage.com}) for English language editing. 
The data presented herein were obtained at the W.M. Keck Observatory, which is operated as a scientific partnership among the California Institute of Technology, the University of California and the National Aeronautics and Space Administration (NASA). The Observatory was made possible by the generous financial support of the W.M. Keck Foundation. 
Some of our observations with the Keck/DEIMOS were carried out under the time-exchange program between the Subaru and the Keck telescopes. The Subaru telescope is operated by the National Astronomical Observatory of Japan. 
The authors wish to recognize and acknowledge the very significant cultural role and reverence that the summit of Maunakea has always had within the indigenous Hawaiian community. 

%

\vspace{5mm}
\facilities{Keck:II (DEIMOS), Subaru (NAOJ)}


\software{SExtractor \citep{BertinArnouts96}, 
             IRAF \citep{Tody86,Tody93}, 
             HYPERZ \citep{Bolzonella+00}, 
             astropy \citep{Astropy+13,Astropy+18}, 
             SDFRED \citep{Yagi+02,Ouchi+04}, \\
             MOPEX (\url{https://irsa.ipac.caltech.edu/data/SPITZER/docs/dataanalysistools/tools/mopex/}), \\
             DSIMULATOR (\url{https://www2.keck.hawaii.edu/inst/deimos/dsim.html}), 
             spec2d \citep{Cooper+12,Newman+13}, 
             specpro \citep{Masters+11}
          }



\appendix

\section{Summary figures for spectroscopically confirmed objects in SSA22-HIT}\label{sec:ap_HITspec}

\begin{figure}
\figurenum{17}
\plotone{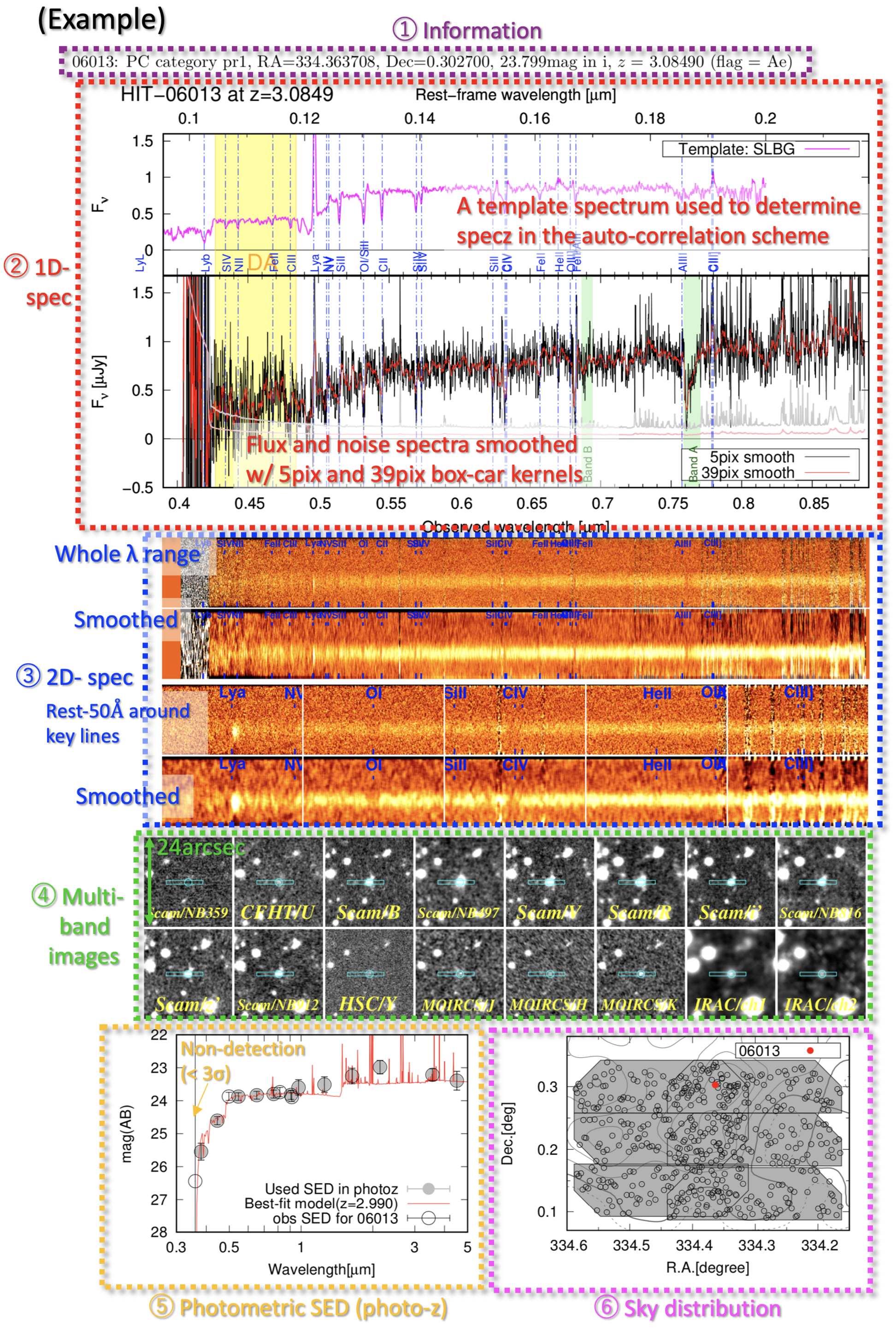}
\caption{
Summary figures for every individual objects whose redshifts were confirmed by the SSA22-HIT survey.
This is an example image using HIT-06013 with descriptions.
There are six types of information in each figure:
1) basic information (e.g., ID\_HIT, R.A., decl., and redshift; see also Table~\ref{aptb:HITcat}) of the given object,
2) one-dimensional spectra with a spectral template used to determine the spectroscopic redshift (Section~\ref{sec:zid_HIT}),
3) two-dimensional spectra,
4) multiband images with the DEIMOS spectroscopic slit (cyan rectangle),
5) an observed SED with the best-fit template that was used in the target selection (Section~\ref{sec:target_selection}),
and 6) sky position of the given object.
The complete figure set (198 images) is available.\label{fig:summary4HIT}}
\end{figure}

In Figure~\ref{fig:summary4HIT}, we show the spectral and imaging data of the 198 spectroscopically confirmed objects in the SSA22-HIT survey (Section~\ref{sec:zid_HIT}). The same figure set is also available via our website\footnote{\url{https://www.icrr.u-tokyo.ac.jp/~mawatari/HIT/PDR/}}. There are six types of information for each object. First, at the top of the figure, the name, category, target priority, R.A., decl., magnitude, spectroscopic redshift, and redshift quality flag (see Section~\ref{sec:ap_HITcat}) are listed. Next, we show the observed one-dimensional spectra as well as the template spectrum used to determine the redshift (Section~\ref{sec:zid_HIT}). The observed spectra are smoothed with 5 and 39 pixel box-car kernels, which are shown by the black and red lines, respectively. The associated error spectra are also shown by the gray and light-red lines. The vertical dotted-dashed lines represent the wavelengths of the possible emission/absorption lines. The yellow and green shaded regions correspond to the wavelength range used to analyze the foreground H\,{\sc i} Ly$\alpha$ absorption and the atmospheric absorption bands, respectively. Third, below the one-dimensional spectra, the original and smoothed two-dimensional spectral images are shown. We show not only the whole wavelength range but also the rest-frame 50\,\AA\ ranges around the Ly$\alpha$, O\,{\sc i} (1302.17\,\AA\ in the rest-frame), C\,{\sc iv} (1548.19\,\AA\ and 1550.77\,\AA), He\,{\sc ii} (1640.35\,\AA), and C\,{\sc iii}] (1906.68\,\AA\ and 1908.73\,\AA). Fourth, the multiband stamp images are shown, where the panel size is $24$\arcsec $\times 24$\arcsec. In each panel, the superposed cyan circle and rectangle represent the 2\arcsec diameter aperture put on the given object and the assigned DEIMOS slit, respectively. Fifth, at the bottom left of the figure, the spectral energy distribution (SED) from the multiband photometry is shown. The observed magnitudes are shown by the circles, where the filled (open) circles correspond to the measurements used (not used) for the SED fitting. The error bar extending across the entire panel represents nondetection ($< 3\sigma$) in the given band. The red spectrum corresponds to the best-fit template in the SED fitting (Section~\ref{sec:target_selection}). Finally, at the bottom right of the figure, the sky position of the given object is shown by the red filled circle. The black open circles, gray shaded regions, and contours correspond to all of the targets in the DEIMOS observations, the six DEIMOS masks, and the $z = 3.1$ LAE number density contours, respectively (the same as those in Figure~\ref{fig:skydist_targets}).

\figsetstart
\figsetnum{17}
\figsettitle{Summary figures}

\figsetend

\section{SSA22-HIT catalog}\label{sec:ap_HITcat}

The properties of all of the observed objects in the SSA22-HIT observations are published in Table~\ref{aptb:HITcat} machine-readable format. A portion is shown here for guidance regarding its form and content. This catalog is called ``the SSA22-HIT catalog'' and also available via our website (see footnote 9).

In the SSA22-HIT catalog, the position-based names (column~1) are constructed using the selected IAU prefix ``SSA22HIT'' plus JHHMMSS.S+DDMMSS. The ID\_HIT identifiers (column~2) are combinations of the mask ID (the first two digits; see Section~\ref{sec:target_selection}) and the slit ID (the following three digits). The R.A. (column~3) and decl. (column~4) coordinates are measured under our astrometric calibration (Section~\ref{sec:imaging_data}). The SSA22-HIT catalog also contains the observed magnitudes (column~5; Section~\ref{sec:imaging_data}) in specific bands (column~6). 

Column~7 represents the target category described in section~\ref{sec:target_selection}. The T2/T3 category objects are divided into those with and without previous spectroscopic redshift measurements, which are referred to as T2prevz/T3prevz and T2/T3 in the SSA22-HIT catalog. 

Column~8 is the target selection priority in each category adopted in the DEIMOS slit assignment using DSIMULATOR (Section~\ref{sec:target_selection}). For the five major categories (T1, T2, T3, T4, and T5), the priorities are set according to whether the objects satisfy the color criteria and whether their photo-$z$ measurements satisfy the expected redshift range. The highest priority (pr1) is set to the objects satisfying both the color criteria and the photo-$z$ range. The second and third priorities (pr2 and pr3) are set to the objects satisfying only the color criteria and only the photo-$z$ range, respectively. We also set the lowest priority (pr4) for objects satisfying only slightly relaxed color criteria. The LyC category objects are prioritized in the following order: spectroscopically confirmed from previous observations (pr1), $i'$-band detected (pr2), and $i'$-band not detected (pr3). For the SMG category, the highest priority (pr1) is set to the SMGs whose SEDs are likely to be at $z \sim 5$, and the second and third priorities (pr2 and pr3) are set to the bright and faint SMGs in the $z'$-band, respectively. The priorities for the LDRLAE category objects are set according to their Ly$\alpha$ emission line rest-frame equivalent width (EW$_0$) photometrically estimated from the $BV -NB497$ color: EW$_0 \gtrsim 200$\,\AA\ (pr1), $45$\,\AA\ $\lesssim$ EW$_0 \lesssim 200$\,\AA\ (pr2), and $15$\,\AA\ $\lesssim$ EW$_0 \lesssim 45$\,\AA\ (pr3). For the galDLA category, we set the highest priority (pr1) for the sight-line galaxy itself and the second priority (pr2) for the candidate counterparts. There is no priority difference in the T2specz, T3specz, oldLAE, and PC targets. 

Column~9 represents the spectroscopic redshifts determined using the specpro software (\citealt{Masters+11}; Section~\ref{sec:zid_HIT}). The quality flags and the best-fit templates in the redshift determination are given in column~10 and 11, respectively. 

\begin{table}[tbp]
\begin{center}
\caption{SSA22-HIT Catalog} \label{aptb:HITcat}
\begin{scriptsize}
\begin{tabular}{lcccccccccc}
\hline
\hline
Name & ID\_HIT & R.A. & Decl. & Magnitude & Band & Category & Priority & Redshift & Flag & Template \\
  &  & (deg) & (deg) &  &  &  &  &  &  &  \\
\hline
{\scriptsize SSA22HIT J221717.3+001009} & 01006 & 334.322000 & 0.169256 & 24.977 & i & T1 & pr3 & -99 & None & None \\
{\scriptsize SSA22HIT J221820.2+000939} & 01007 & 334.584333 & 0.160956 & 24.726 & i & T1 & pr3 & -99 & None & None \\
{\scriptsize SSA22HIT J221729.9+000749} & 01008 & 334.374667 & 0.130311 & 24.675 & i & T1 & pr3 & 2.76676 & Ae & SLBG \\
{\scriptsize SSA22HIT J221752.0+000745} & 01009 & 334.466875 & 0.129042 & 25.495 & i & T1 & pr1 & -99 & None & None \\
{\scriptsize SSA22HIT J221800.4+000553} & 01010 & 334.501542 & 0.098161 & 25.044 & i & T1 & pr1 & -99 & None & None \\
{\scriptsize SSA22HIT J221757.5+000645} & 01011 & 334.489417 & 0.112364 & 24.164 & i & T1 & pr1 & 2.77246 & Aa & HITdr1aLBG \\
{\scriptsize SSA22HIT J221743.7+000916} & 01012 & 334.432167 & 0.154558 & 26.504 & NB359 & LyC & pr3 & -99 & None & None \\
{\scriptsize SSA22HIT J221814.5+001016} & 01013 & 334.560500 & 0.171089 & 25.624 & z & T4 & pr3 & -99 & None & None \\
\hline
\end{tabular}
\end{scriptsize}
\end{center}
\tablecomments{The IAU prefix selected for the identifiers in this catalog is SSA22HIT plus JHHMMSS.S+DDMMSS. This table is published in its entirety in the machine-readable format. A portion is shown here for guidance regarding its form and content.}
\end{table}

\section{Compiled spectroscopic redshift catalog in the SSA22-Sb1 field}\label{sec:ap_pubcat}

Two types of spectroscopic redshift catalogs, both of which contain not only the SSA22-HIT objects but also previously identified sources, are available in machine-readable format: one is ``the representative redshift catalog'' excluding duplication among the original references (Table \ref{aptb:zcat}), and  the other is ``the complete catalog'' with all available redshift information (Table \ref{aptb:col_compzcat}). Both of the catalogs are also available via our website (see footnote 9). 

In the representative redshift catalog, names (column~1), redshifts (column~4), and object type (column~7) are extracted from the original references (column~6), where we applied the extra redshift calibration to some objects (Section~\ref{sec:otherdata}). We set flags for redshift reliability (column~5) based on the original references. The R.A. (column~2) and Decl. (column~3) coordinates are measured under our astrometric calibration (Section~\ref{sec:imaging_data} and ~\ref{sec:compcat}). AGN types (column~8) are determined by matching with the published catalogs or based on visual inspection of the rest-UV emission lines in the SSA22-HIT spectra (Section~\ref{sec:compcat}). Column~9 represents the reference for the AGN types. Because the complete catalog contains as many as 87 columns, individual column descriptions for the complete catalog are presented in Table~\ref{aptb:col_compzcat} instead of a severely truncated sample.

\begin{table}[tbp]
\begin{center}
\caption{Representative Redshift Catalog} \label{aptb:zcat}
\begin{scriptsize}
\begin{tabular}{lcccccccc}
\hline
\hline
Name & R.A. & Decl. & Redshift & Quality\_Flag & Reference & Object\_Type & AGN\_Type & AGN\_Reference \\
  & (degree) & (degree) &  &  &  &  &  &  \\
\hline
{[HIK2019]} 36016            & 334.304194 & 0.117303 & 2.9510 & A & VIMOS08 & LBG     & NONE   & NONE \\
{[HIK2019]} 33559            & 334.273040 & 0.108492 & 2.9510 & A & VIMOS08 & LBG     & NONE   & NONE \\
{[HIK2019]} 61633            & 334.269873 & 0.204618 & 0.2460 & A & VIMOS08 & contami & NONE   & NONE \\
{[HIK2019]} 144725           & 334.302492 & 0.484101 & 3.1100 & A & VIMOS08 & AGN     & OptAGN & VIMOS08 \\
{[HIK2019]} 133686           & 334.278146 & 0.444815 & 3.1320 & A & VIMOS08 & AGN     & AGN1   & Micheva+17 \\
{[HIK2019]} 48642            & 334.270346 & 0.161246 & 2.4200 & A & VIMOS08 & AGN     & OptAGN & VIMOS08 \\
SSA22VI12 J221651.4+001832 & 334.214124 & 0.308907 & 0.3315 & A & VIMOS12 & contami & NONE   & NONE \\
SSA22VI12 J221652.7+002417 & 334.219563 & 0.404811 & 3.0650 & B & VIMOS12 & LyC     & NONE   & NONE \\
\hline
\end{tabular}
\end{scriptsize}
\end{center}
\tablecomments{The literature identifiers used in this table are
    reproduced to (a) match as closely as possible the identifiers
    used in the original references and (b) be formatted to match
    naming aliases used by NED and Simbad. For sources from
    previously unpublished catalogs, we created new position-based
    identifiers based upon an IAU-normalized prefix consisting of
    "SSA22" + an identifying 3-4 digit code + JHHMMSS.s+DDMMSS, e.g.,
    SSA22M05 for sources from the previously unpublished
    \citet{Matsuda+05} data. (This table is available in its entirety
    in machine-readable form.) }
\end{table}


\begin{table}[tbp]
\begin{center}
\caption{Column Description for the Complete Catalog} \label{aptb:col_compzcat}
\begin{scriptsize}
\begin{tabular}{ccl}
\hline
\hline
Column \# & ID & Description \\
\hline
1 & Name                 & Identifiers in the references yielding the representative redshifts \\
2 & RAdeg                & R.A. in units of degrees, measured under our astrometric calibration (Section~\ref{sec:imaging_data}) \\
3 & DEdeg                & decl. in the units of degrees, measured under our astrometric calibration (Section~\ref{sec:imaging_data}) \\
4 & r\_zspec             & References yielding the representative redshifts (Section~\ref{sec:compcat}) \\
5 & Ndup                 & Number of duplicated references for the given object \\
6 & Name-Erb+14          & Identifiers in \citet{Erb+14} \\
7 & z-Erb+14             & Redshifts in \citet{Erb+14} \\
8 & Flag-Erb+14          & Quality flags in \citet{Erb+14}: ``A'' is higher and ``B'' is lower reliability. \\
9 & ObjType-Erb+14       & Object types in \citet{Erb+14} \\
10 -- 13 & *-Yamanaka+20 & Same as columns 6 -- 9 but in \citet{Yamanaka+20} \\
14 -- 17 & *-ADF22       & Same as columns 6 -- 9 but in the ADF22 \citep{Umehata+17b,Umehata+18} \\
18 -- 21 & *-Kubo+15     & Same as columns 6 -- 9 but in \citet{Kubo+15} \\
22 -- 25 & *-Kubo+16     & Same as columns 6 -- 9 but in \citet{Kubo+16} \\
26 -- 29 & *-Chapman+04  & Same as columns 6 -- 9 but in \citet{Chapman+04} \\
30 -- 33 & *-Chapman+05  & Same as columns 6 -- 9 but in \citet{Chapman+05} \\
34 -- 37 & *-Steidel+03  & Same as columns 6 -- 9 but in \citet{Steidel+03} \\
38 -- 41 & *-Nestor+13   & Same as columns 6 -- 9 but in \citet{Nestor+13} \\
42 -- 45 & *-HIT         & Same as columns 6 -- 9 but in the SSA22-HIT (this work) \\
46 -- 49 & *-VIMOS08     & Same as columns 6 -- 9 but in the VIMOS08 \citep{Hayashino+19} \\
50 -- 53 & *-VIMOS12     & Same as columns 6 -- 9 but in the VIMOS12 (H. Umehata et al. 2023,in preparation) \\
54 -- 57 & *-VIMOS06     & Same as columns 6 -- 9 but in the VIMOS06 \citep{Kousai11} \\
58 -- 61 & *-Matsuda+06  & Same as columns 6 -- 9 but in \citet{Matsuda+06} \\
62 -- 65 & *-Matsuda+05  & Same as columns 6 -- 9 but in \citet{Matsuda+05} \\
66 -- 69 & *-Yamada+12   & Same as columns 6 -- 9 but in \citet{Yamada+12b} \\
70 -- 73 & *-DEIMOS08    & Same as columns 6 -- 9 but in the DEIMOS08 \\
74 -- 77 & *-VVDS        & Same as columns 6 -- 9 but in the VVDS \citep{Garilli+08,LeFevre+13} \\
78 -- 81 & *-Saez+15     & Same as columns 6 -- 9 but in \citet{Saez+15} \\
82 -- 85 & *-oIMACS      & Same as columns 6 -- 9 but in the oIMACS \\
86 & AGN                 & AGN types (Section~\ref{sec:otherdata}) \\
87 & r\_AGN              & References for AGN types (Section~\ref{sec:otherdata}) \\
\hline
\end{tabular}
\end{scriptsize}
\end{center}
\tablecomments{The literature identifiers used in this table are
    reproduced to (a) match as closely as possible the identifiers
    used in the original references and (b) be formatted to match
    naming aliases used by NED and Simbad. For sources from
    previously unpublished catalogs, we created a new position-based
    identifiers based upon an IAU-normalized prefix consisting of
    ``SSA22'' + an identifying 3-4 digit code + JHHMMSS.s+DDMMSS, e.g.,
    SSA22M05 for sources from the previously unpublished
    \citet{Matsuda+05} data. (This table is available in its entirety
    in machine-readable form.) }
\end{table}




\bibliographystyle{aasjournal}
\bibliography{HITsurvey_draft}

\end{document}